# Assessing the maturity of software testing services using CMMI-SVC: An industrial case study

Vahid Garousi
Queen's University Belfast, UK
v.garousi@qub.ac.uk

Seyfettin Arkan
Samar IT A.Ş., Adana, Turkey
seyfifurkan@gmail.com

Gökhan Urul
Kolt IT A.Ş., Ankara, Turkey
gokhanurul@gmail.com

Çağrı Murat Karapıçak
Kuasoft IT A.Ş., Ankara, Turkey
cmkarapicak@kuasoft.com

Michael Felderer
University of Innsbruck, Austria
Blekinge Institute of Technology, Sweden
michael.felderer@uibk.ac.at

**Abstract:**

*Context:* While many companies conduct their software testing activities in-house, many other companies outsource their software testing needs to other firms who act as software testing service providers. As a result, *Testing as a Service* (TaaS) has emerged as a strong service industry in the last several decades. In the context of software testing services, there could be various challenges (e.g., during the planning and service delivery phases) and, as a result, the quality of testing services is not always as expected.

*Objective:* It is important, for both providers and also customers of testing services, to assess the quality and maturity of test services and subsequently improve them.

*Method:* Motivated by a real industrial need in the context of several testing service providers, to assess the maturity of their software testing services, we chose the existing *CMMI for Services* maturity model (CMMI-SVC), and conducted a case study using it in the context of two Turkish testing service providers.

*Results:* The case-study results show that maturity appraisal of testing services using CMMI-SVC was helpful for both companies and their test management teams by enabling them objectively assess the maturity of their testing services and also by pinpointing potential improvement areas.

*Conclusion:* We empirically observed that, after some minor customization, CMMI-SVC is indeed a suitable model for maturity appraisal of testing services.

## Table of Contents

## 1 Introduction

Software testing is an impotent phase of the software development life-cycle. At the same time, testing is a costly activity. A 2013 study by the Cambridge University [1] states that the global cost of detecting and fixing software defects has risen to $312 billion annually and it makes up half of the development time of the average project. The most ımportant leverage point for cost-effective improvement of testing is improvement of the entire software development process.

According to various studies, e.g., [2-4], software testing practices and processes in many companies are far from being mature and are usually conducted in ad-hoc fashions. Such immature practices lead to various negative outcomes, e.g., ineffectiveness of testing practices in detecting all the defects, and cost and schedule overruns of testing activities. Also, testing is often conducted not efficiently, e.g., Taipale and Smolander reported that [5]: "*The costs of testing of a software project or product are considerable and therefore it is important to identify process improvement propositions for testing*".

While many companies conduct their software testing activities in-house, there are many other companies who outsource their software testing needs to other firms, usually dedicated to providing software testing services [2-4, 6-9]. Outsourcing enables an organization to concentrate on software development activities while external software testing experts handle the independent validation work. This offers many business benefits which include independent appraisal leading to enhanced delivery confidence, reduced time to market, lower infrastructure investment, predictable software quality, de-risking of deadlines and increased time to focus on development [10-12].

In the context of software testing services and in the course of delivering such services, there could be various challenges and, thus as a result, the quality of test services is not always as expected by service clients. Thus, it is important, for both providers and also customers (clients) of test services, to assess the quality and maturity of test services and subsequently improve them.

Test Maturity Assessment/appraisal (TMA)[1] and Test Process Improvement (TPI) are two active areas among both researchers and practitioners. To improve the quality of technical software testing activities (e.g., test-case design, usage of test metrics), various TMA and TPI models and approaches been proposed by practitioners and researchers, e.g., Test Maturity Model integration (TMMi) [13, 14] and the Test Process Improvement (TPI) model and its successor TPI-Next [15].

A recent Multivocal Literature Review (MLR) study, in which the first author was involved, was conducted on the subject of TMA and TPI [16]. Let us recall from [16] that a MLR is a form of a Systematic Literature Review (SLR) which includes the grey literature (e.g., blog posts and white papers) in addition to the published (formal) literature (e.g., journal and conference papers). The MLR systematically selected and reviewed 181 studies on the topic, and developed a list of all 58 test maturity models, proposed so far by practitioners and researchers. Another recent 2016 SLR [17] on this topic identified 18 TPI approaches showing the fast progress of this important field in software testing.

In the course of many (150+) industrial software testing projects, e.g., [18-22] (both ongoing and also in the past) for the case of all four authors, they have observed that many companies have the need for assessing maturity of software testing services. Although there are various established models such as TMMi [14] to assess and improve maturity of technical testing tasks (e.g., test-case design and test automation), but as of this writing, no study or model has been proposed for maturity appraisal of software testing 'services'. We have observed situations in which a testing service provider has done an effective job in technical testing tasks, but has failed to provide high quality of testing 'service' w.r.t. its service obligations, e.g., failure to meet service delivery objectives, and ineffective relationship with test sub-contractors. Many studies, e.g., [23, 24], have argued that guidance on developing and improving mature service practices is a key contributor to the service provider performance and customer satisfaction. This is also the case for testing services.

In the context of software testing service projects, provided / received by the authors and their partners, we have faced a need to assess the maturity of software testing services, from the viewpoint of both service providers and service customers. To address that real need, and based on an the principles of "action research" [25, 26], and a technology transfer model proposed by Gorschek et al. [27], we planned and conducted an empirical study which we report in this paper. The industry-academia collaboration (IAC) was planned and executed using the best practices in the literature to ensure success in IACs in software engineering [28] and based on our past experience in conducting IACs in software testing [19].

Instead of developing another yet new maturity model, given the large number of existing maturity models in software engineering [16, 29], we searched for candidate maturity models which could be possibly applied for assessing maturity of testing service, e.g., the 'CMMI for Services' (CMMI-SVC) model [23, 24], ISO/IEC 20000 [30] and the Information Technology Infrastructure Library (ITIL) [31]. After a systematic comparison which re report in this paper, we chose the CMMI-SVC, and customized it for our purpose.

The remainder of this paper is structured as follows. A review of the background and related work is presented in Section 2. We formally characterize test services in Section 3. In Section 4, we discuss how we customized CMMI-SVC for test services. Section 5 presents a case study in which the approach is applied to two companies providing software testing services. Finally, in Section 6, we draw conclusions, and suggest areas for further work.

---

[1] A summary of the acronyms used in the paper is provided in the appendix.

## 2 BACKGROUND AND RELATED WORK

We review in this section the following topics:

- Emergence of software testing as a service (Section 2.1)
- State-of-the-art in test maturity assessment/appraisal and test process improvement (Section 2.2)
- A review of the relevant service maturity models (Section 2.3)
- Relationship among "test"-focused and "service" -focused maturity models (Section 2.4)
- Choosing a base maturity model for our needs (Section 2.5)

### 2.1 EMERGENCE OF SOFTWARE TESTING AS A SERVICE

Many companies outsource their software testing needs to other firms who are usually dedicated to providing software testing services. There are various reasons to outsource testing, e.g., lack of in-house test experts, reducing test costs, and strict delivery deadlines.

Software testing are outsourced in different forms [10-12]: (1) Full outsourcing, insourcing or remote insourcing of the entire test process (strategy, planning, execution and closure), often referred to as a Managed Testing Service or dedicated testing teams; (2) Provision of additional resources for major projects; (3) One-off test often related to load, stress or performance testing; and (4) Beta User Acceptance Testing. Utilising specialist focus groups coordinated by an external organization.

As a a result, *Testing as a Service* (TaaS) has emerged as a strong service industry in the last several decades, e.g., [10-12]. The size of this market worldwide has been reported to be in billions annually [32]. A market research report [32] by Research and Markets Co. announced in 2015 that the global Testing-as-a-Service market will grow at the rate of 11% during the forecasted period for 2016-2020.

Testing-as-a-Service has become a popular topic in the grey literature and among practitioners, e.g., a Google search for "*software testing services*" as of this writing (Mar. 5, 2017) returned 380,000 hits. There are many weblogs and industrial case studies (often in the form of white papers) on the topic. For example, a large firm named Infosys Limited has published summaries of their industrial testing service projects as case study white papers in [33].

There has also been research on various aspects of Testing-as-a-Service, e.g., [7-11, 34]. The work in [7] reported a case study to assess the challenges of managing Testing-as-a-Service. The paper in [8] studied client communication practices in managing relationships with offshore vendors of Testing-as-a-Service. Vendor-side experiences in Testing-as-a-Service were studied in [9]. The work in [10] presented a classification of different types of testing services in Testing-as-a-Service, and provided a comparative view and perspectives between conventional software testing service and cloud-based Testing-as-a-Service. In addition, it examined underlying issues, challenges, and emergent needs. The work in [11] proposed a reference architecture of Testing-as-a-Service based on ontology, process automation and SOA techniques. The work in [34] proposed a dynamic life-cycle model for the provisioning of Testing-as-a-Service and reported the experiences from a case study in the Chinese software market.

We will discuss the context and processes for establishment and delivery of test services in detail in Section 3. Since providing test services includes both technical testing tasks (e.g., test-case design and test execution) and service-related tasks (e.g., coordination with the client and test tool vendors, and managing changes in test requirements and software requirements), testing important is in service perspective and there is a need to study maturity of test services.

### 2.2 A REVIEW OF TEST MATURITY MODELS AND A REVIEW OF TMMI AND TPI

As discussed in Section 1, TMA and TPI are active areas among both researchers and practitioners. A recent Multivocal Literature Review (MLR) in this area [16, 35] systematically selected and reviewed 181 studies, and identified 58 different test maturity models. Also another recent 2016 SLR [17] on this topic identified 18 TPI approaches showing the fast progress of this important field in software testing.

Due to space constraints, we do not re-list all the 58 test maturity models identified in [16], again in this article, but instead, we only present a few examples in Table 1, while the full list can be found in [16]. In terms of popularity (number of usages in the studies), the MLR [16] found that TMMi (and its earlier version TMM) [13, 14] and TPI (and its successor TPI-Next) [15] are the most popular models. We provide a brief overview of TMMi and TPI in the following.

**Table 1-Examples of the test maturity models proposed in the community along with their maturity levels (taken from [16])**

| | | |
|---|---|---|
| Test Maturity Model integration (TMMi) [13, 14]: a 'staged' model<br>• Level 1: Initial<br>• Level 2: Definition<br>• Level 3: Integration<br>• Level 4: Management and measurement<br>• Level 5: Optimization | Test Process Improvement (TPI) [15]: a 'continuous' model, i.e., not 'staged' (based on maturity levels), but including 20 Key Performance Areas (KPAs). Each KPA has four levels: A...D<br>1. Test strategy<br>2. Life-cycle model<br>3. Moment of involvement<br>….<br>18. Test process management<br>19. Evaluation<br>20. Low-level testing | Unit Test Maturity Model [36]<br>• Level 0: Ignorance<br>• Level 1: Few simple tests<br>• Level 2: Mocks and stubs<br>• Level 3: Design for testability<br>• Level 4: Test driven development<br>• Level 5: Code coverage<br>• Level 6: Unit tests in the Build<br>• Level 7: Code coverage feedback Loop<br>• Level 8: Automated builds and tasks |
| Agile Quality Assurance Model (AQAM) [37]<br>• Level 1: Initial<br>• Level 2: Performed<br>• Level 3: Managed<br>• Level 4: Optimized | Automated Software Testing Maturity Model (ASTMM) [38]<br>• Level 1: Accidental automation<br>• Level 2: Beginning automation<br>• Level 3: Intentional automation<br>• Level 4: Advanced automation | TPI-EI [39]<br><br>Adaptation of TPI for embedded software |
| Agile Testing Maturity Model (ATMM) [40]<br>• Level 0: Waterfall<br>• Level 1: Forming<br>• Level 2: Agile bonding<br>• Level 3: Performing<br>• Level 4: Scaling | TestSPICE [41]<br><br>A set of KPAs. Based on ISO/IEC 15504, Software Process Improvement and Capability dEtermination (SPICE) standard | The Personal Test Maturity Matrix [42]<br><br>A set of KPAs such as: test execution, automated test support and reviewing |

TMMi is based on the Capability Maturity Model (CMM) and CMMI models, and its first version was proposed in 1998 [43]. The latest version of TMMi specification as of this writing is 1.0 [14] prepared and published by the TMMi Foundation in 2012.

Figure 1 shows the maturity levels and process areas of TMMi. As the structure shows, each maturity level has several process areas (PA), and each PA has several Specific Goals (SG). Each SG in turn has several Specific Practices (SP). A PA is a cluster of related practices in an area that, when implemented collectively, satisfies a set of goals considered important for making improvement in that area. A specific goal (SG) describes the unique characteristics that must be present to satisfy the process area. A SG is a required model component and is used in appraisals to help determine whether a process area is satisfied. A specific practice (SP) is the description of an activity that is considered important in achieving the associated specific goal. The SPs describe the activities that are expected to result in achievement of the SGs of a PA. It is a 4-level hierarchical structure. Essentially, there are "aggregation" relationships between a PA and its underlying SGs, and a SG and its underlying SPs (as shown in Figure 1).

The "initial" level # 1 is the one in which test activates are conducted in "ad-hoc" fashion and thus it has no PAs. Under the four maturity levels above the initial level #1 (2, 3, 4 and 5), the TMMi [14] has 50 SGs and 188 SPs, in total. For example, under the level 2 ("Managed"), there are five PAs, e.g., PA 2.1 (test policy and strategy). This PA has three SGs: (SG 1)-establish a test policy, (SG 2)-establish a test strategy, and (SG 3)-establish test performance indicators. The above SG 1, in turn, has three SPs: (SP 1.1)-define test goals, (SP 1.2)-define test policy, and (SP 1.3)-distribute the test policy to stakeholders. Further details can be found in [14].

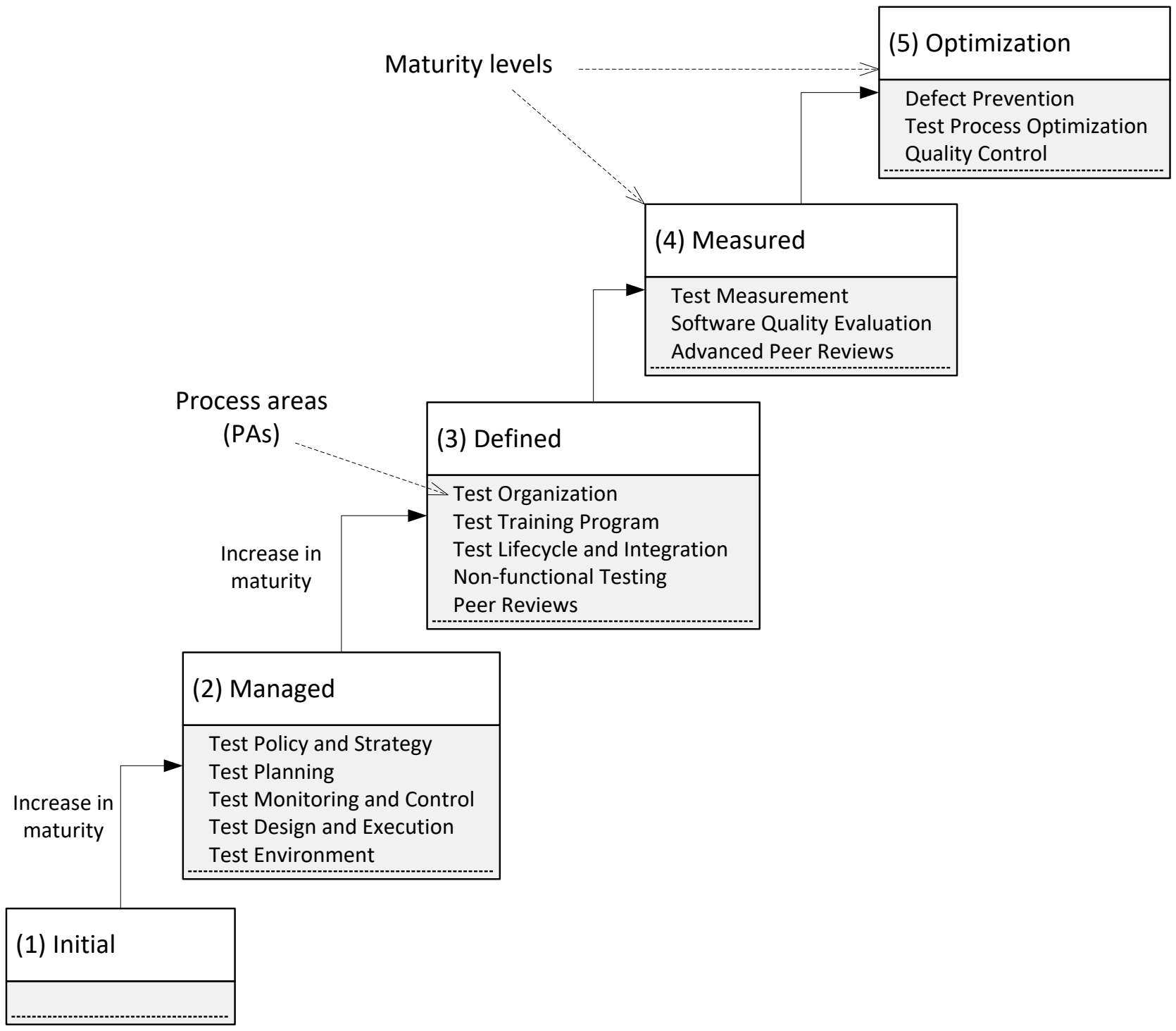

**Figure 1- TMMi maturity levels and process areas [14]**

The Test Process Improvement (TPI) model [15] and its newer version TPI-Next [44] have been developed and maintained by a Nederland-based company named Sogeti since 1998. TPI and TPI-Next are composed of a set of 'Key Areas' (KAs) which are the basis for improving and structuring the test process. Slightly similar to TMMi, TPI differs from TMMi in that it is not staged (maturity-level-based), but instead is continuous and has a set of 20 KAs, e.g., test strategy, and test estimating and planning. Each KA has two to four pre-defined levels which are determined, for the context under study, during maturity assessment. Table 2 shows the list of the levels for five of the TPI's 20 KAs. In the levels of each KA in Table 2, the maturity increases from the levels in the left to the right. For example, for KA#1 (test strategy), there are four levels as shown (A…D). Level A is the least mature level in which there is a "*test strategy for single high-level test*" [15]. Level D is the most mature in which the context under study has "*combined strategy for all test and evaluation levels*".

**Table 2- TPI's maturity levels and several example KAs of the model [15]**

| **Key area** | | **Levels (→increasing maturity from left to right)** | | | |
|---|---|---|---|---|---|
| **#** | **Name** | **A** | **B** | **C** | **D** |
| 1 | Test strategy | Strategy for single high-level test | Combined strategy for high-level tests | Combined strategy for high-level tests plus low-level tests or evaluation | Combined strategy for all test and evaluation levels |
| 2 | Life-cycle model | Planning, specification, execution | Planning, preparation, specification, execution, completion | | |
| 3 | Moment of involvement | Completion of test basis | Start of test basis | Start of requirements definition | Project initiation |
| 4 | Estimating and planning | Substantiated estimating and planning | Statistically substantiated estimating and planning | | |
| … | | | | | |
| 20 | Low-level testing | Low-level test lifecycle: planning, specification and execution | White-box techniques | Low-level test strategy | |

### 2.3 Relevant service maturity models

Since our context in this work is focused on maturity of software testing services, our literature review also covered the maturity models focusing on "services". We were able to find three such models in the literature, which we review next: ISO/IEC 20000 [30], Information Technology Infrastructure Library (ITIL) [31], and CMMI for Services [23, 24].

ISO/IEC 20000 [30] is an international standard for IT service management. It specifies requirements for service providers to plan, establish, implement, operate, monitor, review, maintain and improve a service management system. The requirements include the design, transition, delivery and improvement of services to fulfil agreed service requirements.

The Information Technology Infrastructure Library (ITIL) [31] is a set of practices for IT service management that focuses on aligning IT services with business needs. In its current form (known as ITIL 2011 edition), ITIL is published as a series of five core areas, each of which covers a different IT service management lifecycle stage. The core areas of ITIL include: service strategy, service design, service transition, service operations and continual service improvement.

CMMI for Services (CMMI-SVC) [23, 24] is a member of the well-known CMMI family of maturity models [45]. CMMI is a process improvement training and appraisal program and service, administered and marketed by the Software Engineering Institute (SEI) at Carnegie Mellon University and required by many government contracts in the USA and many other countries, especially in software development. CMMI can be used to guide process improvement across a project, division, or an entire organization. CMMI defines the following maturity levels for processes: (level 1): initial, (level 2) managed, (level 3) defined, (level 4) quantitatively managed, and (level 5) optimizing.

Adapted from [46], Figure 2 shows the relationship among the CMMI family of related maturity models: CMMI for Development (CMMI-DEV), CMMI for Services (CMMI-SVC), and CMMI for Acquisition (CMMI-ACQ). There are two entities in the horizontal axis (supplier/provider and customer) and two entities in the horizontal axis (development and services). The three maturity models cover different aspects w.r.t. these entities.

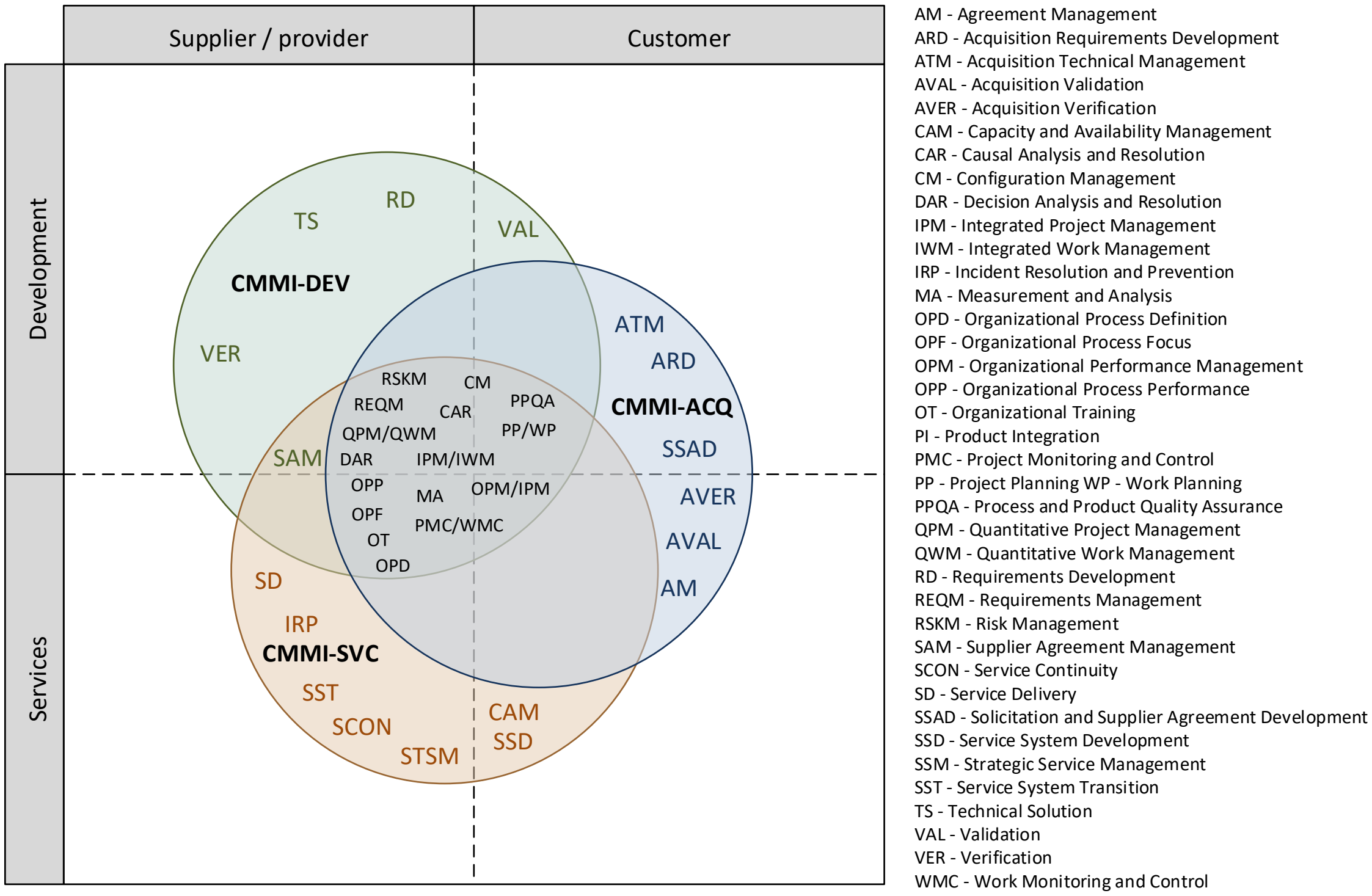

**Figure 2- Relationship among the CMMI family of related maturity models (adopted from: [46])**

As we can see, the major difference between CMMI-DEV and CMMI-SVC is in their Process Areas (PAs). There are in total 38 PAs in Figure 2. Compared to CMMI-DEV, CMMI-SVC excludes engineering (technical) PAs such as RD (Requirements Development) and TS (Technical Solution) while including seven service-related PAs such as SSD (Service System Development) and SST (Service System Transition). CMMI-SVC has four maturity levels (starting from number 2): (level 2) managed, (level 3) defined, (level 4) quantitatively managed, and (level 5) optimizing [23]. Each maturity levels contains several process areas (PAs). The latest version of CMMI-SVC contains 24 PAs in total, 7 of which are specific to services as shown in Figure 2. We are showing in Table 3 the acronyms, full names and descriptions of those seven PAs. These service-specific PAs cover anything about services from service delivery to service continuity and service system development.

**Table 3- Seven service-specific PAs in CMMI-SVC (taken from [23])**

| PA acronym | Full name | Description |
|---|---|---|
| SD | Service delivery | Deliver services in accordance with service agreements put in place in the contract. |
| SCON | Service continuity | The purpose is to establish and maintain plans to ensure continuity of services during and following any significant disruption of normal operations. |
| SSD | Service system development | The purpose is to analyze, design, develop, integrate, verify, and validate service systems, including service system components, to satisfy existing or anticipated service agreements. |
| IRP | Incident resolution and prevention | The purpose is to ensure timely and effective resolution of service incidents and prevention of service incidents as appropriate. |
| CAM | Capacity and Availability Management | The purpose is to ensure effective service system performance and ensure that resources are provided and used effectively to support service requirements. |
| SSM | Strategic service management | Establish and maintain standard services in concert with strategic needs and plans. |
| SST | Service system transition | The purpose is to deploy new or significantly changed service system components while managing their effect on ongoing service delivery. This may include replacing the sub-contractors and practices used in a current service project. |

## 2.4 CHOOSING A BASE MATURITY MODEL FOR THIS STUDY AND OUR CONTEXT

After reviewing "test"-focused and "service" -focused maturity models in Sections 2.2 and 2.3, we now discuss our rationale for choosing a suitable maturity model for this study and our context. As we discussed, test-focused maturity models (such as TMMi and TPI) are focused on "technical" maturity aspects of testing, e.g., maturity of test planning, test-case design and test automation, while service-focused maturity models are focused on delivery, management and continuity of software services. Therefore, for stakeholders who need to assess maturity of testing services, the two aspects (technical maturity and service maturity) shall be assessed separately and, if needed, their results should be carefully combined after separate analysis. It could be that, a service provider may conduct technical testing tasks (e.g., test-case design and test automation) with high maturity, but it may fail to provide effective management of test services (e.g., poor coordination with the client and poor management of test change-requests).

As per the authors' experience in the course of many software testing projects, they have seen the above issues first hand and thus, we carefully separate the two aspects in this work. Various empirical studies on assessing "technical" maturity of testing have been reported (see the list of many studies in [16]), thus we focus in this work on maturity of "service" aspects of testing projects. Therefore, our focus shifts from TMMi and TPI to the models that we reviewed in Section 2.3.

No maturity models have been proposed for assessment of software testing 'services'. Thus, to address our need, we had two alternatives: (1) develop a new model for this purpose from scratch; (2) to choose an existing model and adapt it for our need. After reviewing the above three service maturity models (CMMI-SVC, ISO 20000 and ITIL), we assessed them to see if we can choose one of them for our context.

Sources such as [47] have compared the relationship among the above models. An industry presentation by a company providing testing services [47] reported that: "*CMMI-SVC provides almost complete coverage of ISO 20000 clauses*". The source [47] also believed that , compared to CMMI-SVC, certain practices are not stressed to the same extent in ISO 20000, e.g., stakeholder management. However, stakeholder management is an inherent part of every process area in CMMI-SVC, e.g., what stakeholders are relevant to each task, and how they are involved. Also, [47] reported that CMMI-SVC offers a more detailed set of practices for management of information flows (monitoring and controlling the process), by providing "*pragmatic measures*" to be identified and used, and also a structure for capturing and reporting management information that focuses on what is important to managers.

When comparing CMMI-SVC and ITIL, [47] reported that the structure and content of ITIL is emphatically like a "library" in nature, and it details "how to" implement service practices while CMMI-SVC details "what to" implement and provides a route-map of improvements to implement. In summary, [47] concluded that the three models (CMMI-SVC, ISO 20000 and ITIL) complement each other. While ITIL is a reference library for this purpose, CMMI-SVC is the "reading list" for success and a maturity model, and ISO 20000 is the "exam".

For choosing the base model for our needs, we organized several meetings among the authors who were a mix of industry practitioners and academic researchers. The selection approach was as follows. We were looking for a maturity model and also a model which helps us find the improvement areas. Based on the discussions of [47] and also our own assessments, we found that both ISO 20000 and ITIL are guidelines and not maturity models, per se. Since we were looking for a maturity model, CMMI-SVC was the only option from this standpoint. Both ISO 20000 and ITIL were geared more towards "IT" services, a quite higher level of abstraction compared to "software"-related services. Also, CMMI-SVC provided much better match for contextual and process aspects for establishment and delivery of services, compared to ISO 20000 and ITIL (more details in Section 3). In addition, our industry partners in Turkey were very familiar with CMMI since the Turkish governmental software industry is largely regulated by CMMI. Due to the above reasons, we chose CMMI-SVC as the base model for this work.

## 3 CHARACTERIZING AND FORMALIZING TEST SERVICES

To assess the characteristics and maturity of test services in detail, it is important to properly understand the context and process for establishment and delivery of test services, which we discuss next.

### 3.1 CONTEXT FOR ESTABLISHMENT AND DELIVERY OF TEST SERVICES

In order for us to understand the context and process for establishment and delivery of services, we reviewed the CMMI-SVC specifications [23]. While we found scattered discussions about these issues, we did not find a clear / comprehensive framework in CMMI-SVC for establishment and delivery of services.

Taken from the CMMI-SVC specifications [23], Figure 3 shows how CMMI-SVC defines the "context" for establishing and delivering services. This diagram shows the inter-connection among the various process areas in CMMI-SVC with the service customer (end user).

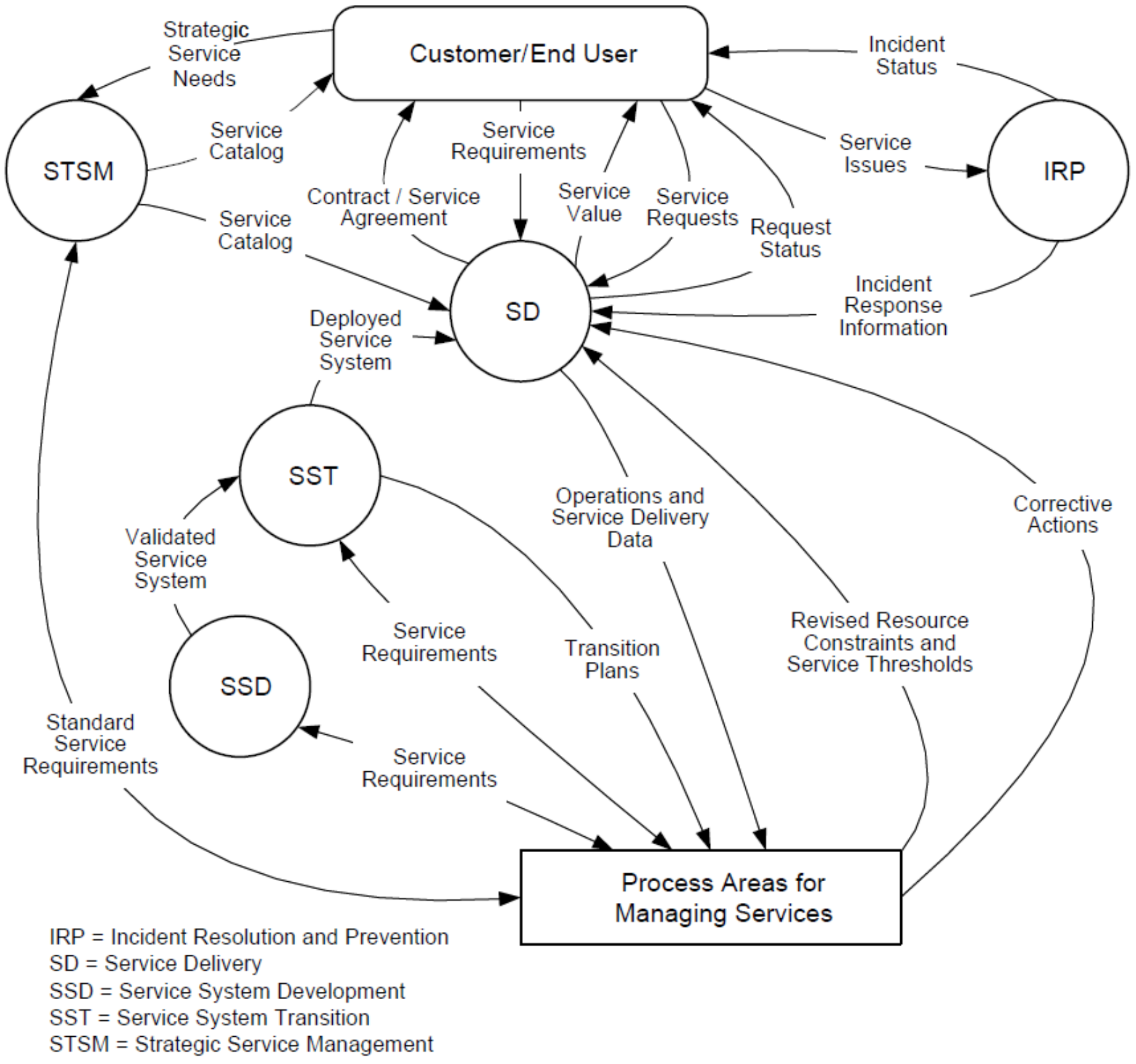

**Figure 3- Context for establishing and delivering services as defined by CMMI-SVC (taken from: [23])**

While the context diagram Figure 3 is a useful model to get an initial understanding of establishment and delivery of test services, in consultation with several industrial test managers, we found that it lacks some important aspects (e.g., the service provider and service sub-contractor actors are excluded), and is also not easy to apply in our partners' industry contexts. It seems that the model was developed in CMMI-SVC as a "rough" explanation of relationships among PAs. We thus adopted some ideas from it (e.g., the notion of service customer) and, by adding our own experience in many test service projects, we developed a revised/improved context diagram for test services as shown in Figure 4.

Our improved context diagram includes five actors: (1) customer (receiver of test services), (2) provider of test services, (3) sub-contractor for test services, (4) supplier of test tools / products, and (5) test staff. While most of the diagram is self-explanatory, we discuss it briefly next.

The focal point of the context diagram is the 'Test Service' entity which connects the two important actors: the customer (receiver), and the provider of service. The scope of the test service is defined via a contract/service agreement which is, usually, signed by the two actors. Provider uses "artifacts needed for testing" to test the Software Under Test (SUT), which customer provides. The test staff of the provider conduct and manage test activities. A test-focused maturity model (such as TMMi) is used to assess the "technical" maturity of test activities, while a service-focused model (such as CMMI-SVC) is used to assess the maturity of test service itself. There is a connection between the two maturity models via process area "Capacity and availability management" of CMMI-SVC in its level #3. "Capacity" in this context refers to technical maturity of test activities.

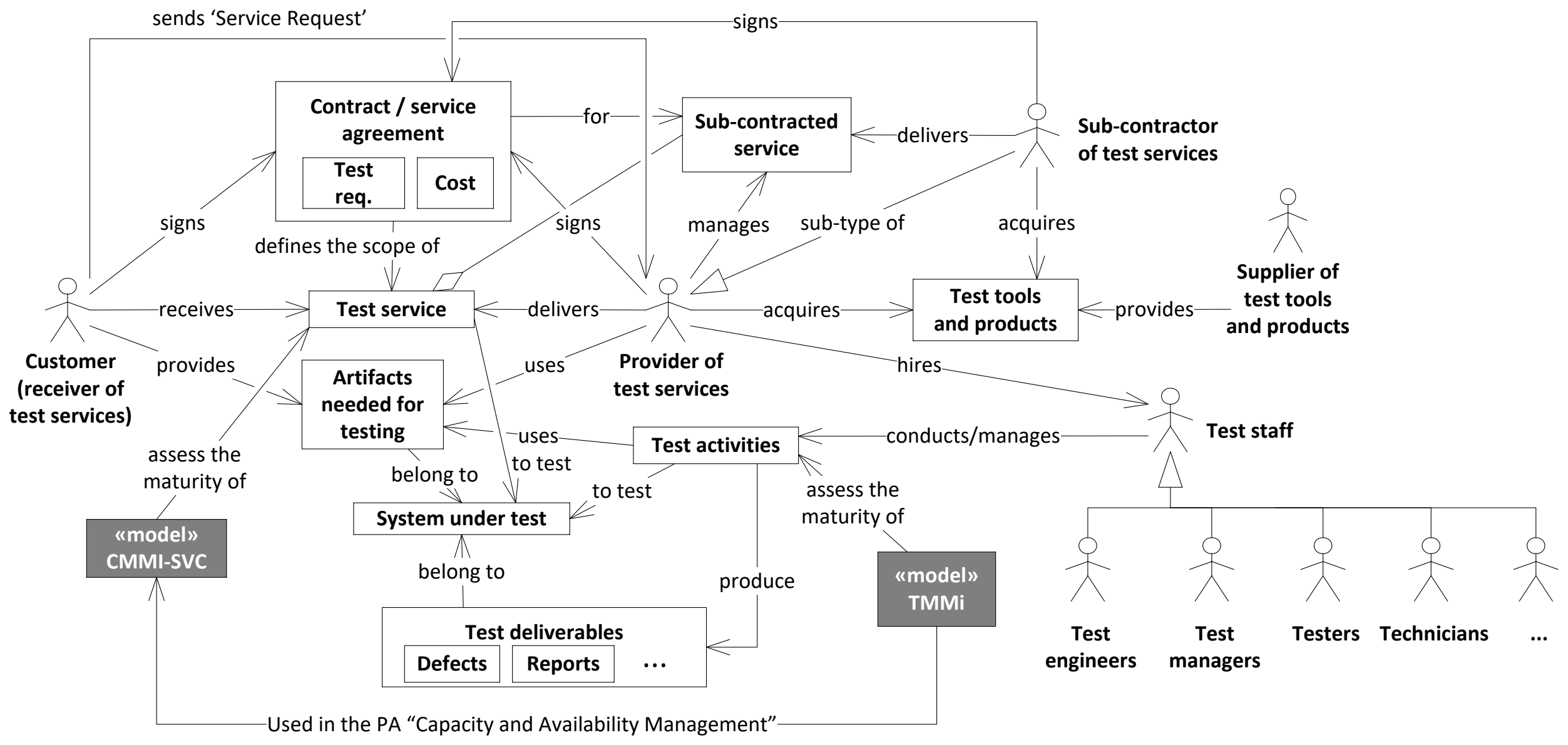

**Figure 4- A UML context diagram for establishment and delivery of test services**

We have incorporated the sub-contractor role as well since, as per our and others experience [48], sub-contracting of test services is quite common in the Testing-as-a-Service industry. If permitted by the contract, a service provider (contractor) may sub-contract parts of the work to sub-contractor(s). A sub-contractor of test services is a type of service provider as shown by the sub-class relationship in Figure 4. When test services are contracted out, the sub-contractor "plays the role of" a provider and the intermediate service provider plays the role of customer (as a proxy).

Finally, if there is a need, test tools and products for carrying out the testing tasks and delivery of test services are acquired. For this purpose, an additional actor (supplier of test tools and products) is considered.

To complement the context model in Figure 4, we discuss next the "process" for establishment and delivery of test services, which is another important aspect in characterizing test services ("context" and "process", in this scope, correspond to structure and behavioral modeling in OO development).

### 3.2 Process for establishment and delivery of test services

By consultation with several of our industry partners and using our industry experience in offering test services, we have formalized the process for establishment and delivery of test services as a UML activity diagram in Figure 5. Similar to Figure 4, there are two primary actors in this process: provider of test services, and customer (receiver of test services). The same process can also apply for a sub-contractor as provider.

The process starts the bidding and contracting phase which itself has these activities in order: (1) specify test service requirements, (2) request for bids, (3) submit bids (bidding), (4) reviewing the bids, (5) awarding the contract, and (6) signing the contract. Once the contract is in place, the planning activity starts. If a sub-contractor (or more) is (are) needed, a sub-process starts to conduct the bidding and contract job involving one or more sub-contractors.

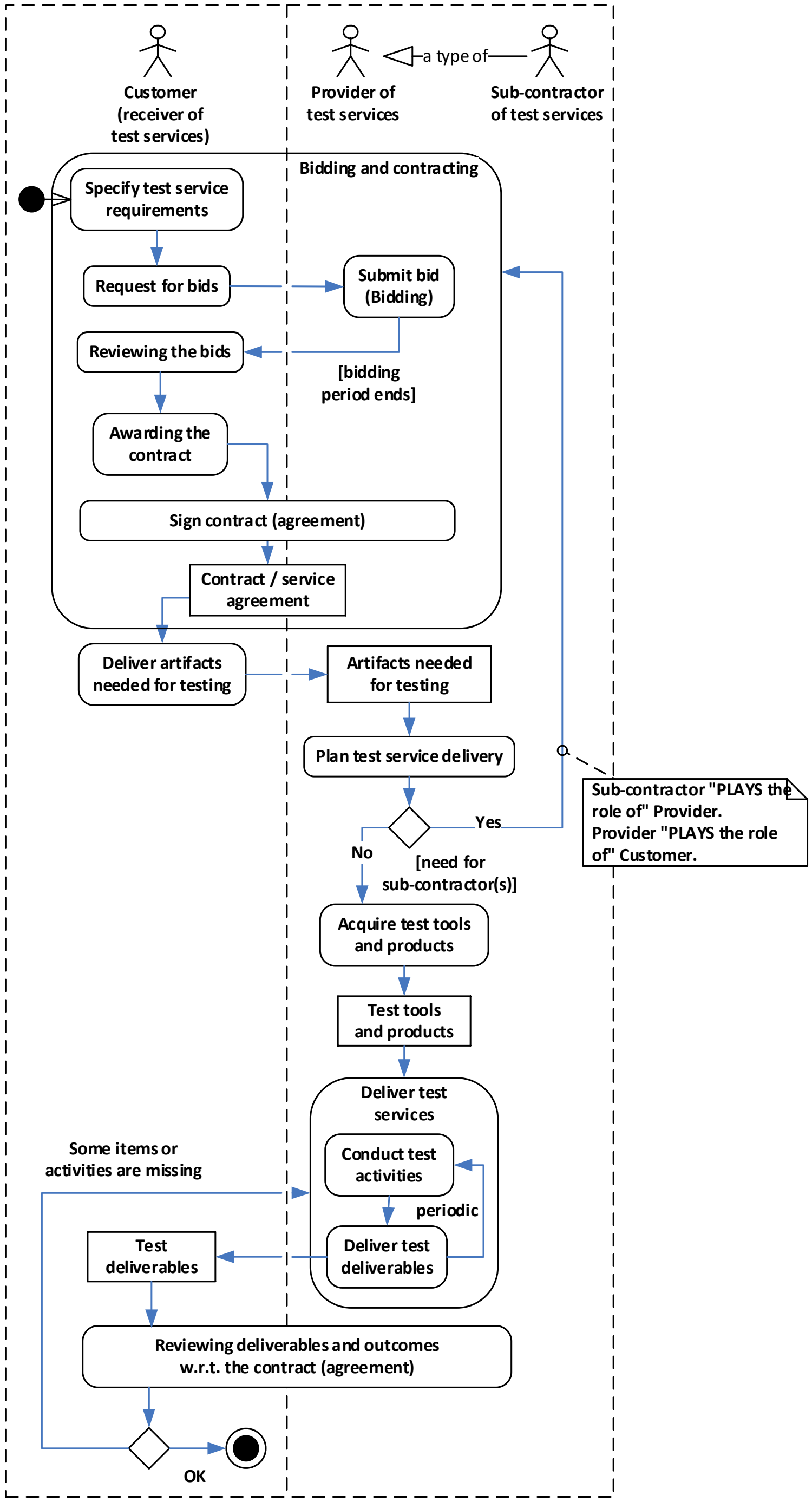

**Figure 5- A UML activity diagram (process) showing the process for establishment and delivery of test services**

Afterwards, if there is a need, test tools and products for carrying out the testing tasks and delivery of test services are acquired. The next activity includes the technical testing tasks, i.e., to deliver the services which includes conducting test activities and delivering test deliverables (e.g., test results, reports and automated scripts). Once the client accepts the quality of the deliverables and the service, the process finishes. The activities and processes associated with service delivery in Figure 8 (i.e., conducting test activities and delivering test deliverables) are considered as "primary processes", which can enable us interpret the entire process (Figure 8) as a set of base practices in primary processes in the ISO perspective [49].

Characterizing and formalizing the context and process for establishment and delivery of test services enables us to better understand what aspects we shall consider in maturity appraisal of software testing services.

## 4 MINOR CUSTOMIZATION OF CMMI-SVC FOR TEST SERVICES

Since CMMI-SVC is a general maturity model for services, we first assessed its applicability for test services by carefully reviewing all of its process areas (PA) and Specific Goals (SG). This was done to ensure that the model would meet the needs of the software testing industry. Two important criteria that we considered while assessing the applicability of CMMI-SVC for test services were its "completeness" and "relevance" in this context.

Completeness of CMMI-SVC for test services denotes whether this model captures all the important aspects in the scope of maturity of testing services, as needed in real-world projects. In other words, from the point of view of stakeholders involved, does the model include all the necessary aspects in testing services? Does the model assess what it aimed at assessing? On the other hand, relevance of CMMI-SVC for test services denotes the extent to which each of the factors and issues included in the model are relevant, i.e., should be assessed in the context of test services. Our applicability assessment was conducted by all four authors of this paper, three of which are working software testing practitioners and regularly provide testing services.

As a result of our applicability assessment, we customized CMMI-SVC slightly for test services. We should note that that we did not intend to develop yet another maturity model (MM) nor to extend CMMI-SVC since our customizations are minor and do not suffice calling it an extension of CMMI-SVC.

Figure 6 shows the maturity levels and process areas of CMMI-SVC and the customizations that we had to make (underlined) after our applicability assessment. Out of the 24 PAs in CMMI-SVC, we revised two PAs and added one new PA. For clarity, preciseness and ease of model's applicability in the Testing-as-a-Service domain, we also reworded and adopted the description of many PAs and SGs to highlight the notion of test services, e.g., "Test Service Delivery" versus "Service Delivery" in CMMI-SVC, or "Test Service Agreement Management" versus "Supplier Agreement Management" in CMMI-SVC.

We also added more details to some of the descriptions to help our industry contacts understand what is asked in the items, e.g., SP 1.2 of PA "Test service agreement establishment and delivery" was revised to include the elements of the test service agreement as follows: "*Discus and negotiate elements of the test service agreement (contract), e.g., basic details such as test requirements, duration/cost, and other issues such as scope of regression tests and when to stop testing*".

For brevity, we do not place the entire details of all model elements in this paper body, but we it can be found in an archived online resource [50]. We discuss next a few important customizations made to CMMI-SVC and their rationale. Other minor customizations and wording refinements can be reviewed in [50].

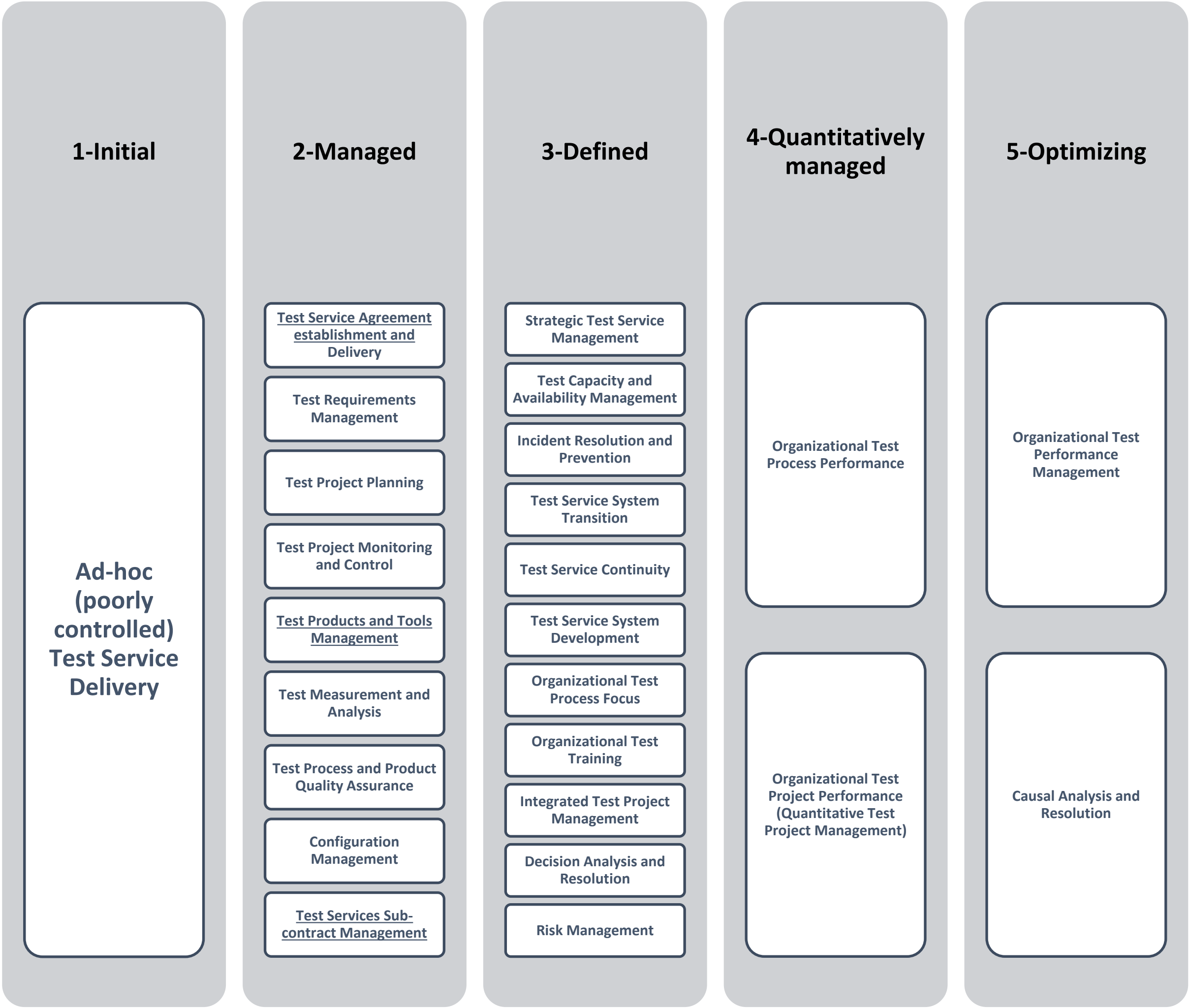

**Figure 6- Model structure of CMMI-SVC and customizations made to it for Testing-as-a-Service (underlined)**

### 4.1 Refining the PA "Test service agreement establishment and delivery"

One of the fundamental PAs in CMMI-SVC is "Service Delivery", defined by the specifications [23] as: "*The purpose of Service Delivery (SD) is to deliver services in accordance with service agreements*". After reviewing the list of its SGs (shown in Table 4), since this PA also includes the notion of "service agreement", we made its naming more precise by rephrasing it to '*Test service agreement establishment and delivery*'. In joint work with practitioners, they informed us that more precise PA titles and descriptions are better to work with and will prevent confusions.

Also as shown in Table 4, we added SP 1.2 under SG 1 of this PA to highlight that discussing and negotiating test service agreements (contracts) are important and shall be included in this PA. Recall the two criteria of completeness and relevance hat we discussed in Section 4.1. The model completeness was improved by this refinement. We also added further details to SP 3.2 and 3.3 by adding the elements of test services (including test staff, test processes and test activities).

**Table 4- Refining the PA "Test service agreement establishment and delivery"**

| Original in CMMI-SVC: | How it was refined: |
|---|---|
| • SG 1 Establish service agreements<br>o SP 1.1 Analyze existing agreements and service data<br>o SP 1.2 Establish the service agreement<br>• SG 2 Prepare for service delivery<br>o SP 2.1 Establish the service delivery approach<br>o SP 2.2 Prepare for service system operations<br>o SP 2.3 Establish a request management system<br>• SG 3 Deliver Services<br>o SP 3.1 Receive and process service requests<br>o SP 3.2 Operate the service system<br>o SP 3.3 Maintain the service system | • SG 1 Establish test service agreements<br>o SP 1.1 Analyze test service and customer needs<br>o SP 1.2 Discuss and negotiate elements of the test service agreement (contract), e.g., basic details such as test requirements, duration/cost, and other issues such as scope of regression testing and when to stop testing<br>o SP 1.3 Establish the test service agreement (contract)<br>• SG 2 Prepare for test service delivery<br>o SP 2.1 Establish the test service delivery approach<br>o SP 2.2 Prepare for test service operations<br>• SG 3 Deliver test services<br>o SP 3.1 Receive and process test service requests<br>o SP 3.2 Operate the test service (including test staff, test processes and test activities)<br>o SP 3.3 Maintain the test service (including test staff, test processes and test activities) |

## 4.2 ADDING THE PA "TEST PRODUCTS AND TOOLS MANAGEMENT"

The notions of products / tools needed for server delivery and their management are quite under-represented in CMMI-SVC. Only under its PA "Supplier agreements management", there is a SP phrased as "Accept the acquired products" which was not too clear in our context. Also we were aware that, similar to the case of in-house testing, test products and tools are very important in the context of Testing-as-a-Service. Thus we added this PA to the model. Table 5 shows the details. This new PA has two SGs and six SPs in total. The rationale of these SGs are to establish requirements for test products and tools, and to manage those products and tools.

**Table 5- Adding the PA "Test products and tools management"**

| Original in CMMI-SVC: | How it was refined: |
|---|---|
| Did not exist in CMMI-SVC | • SG 1 Establish requirements for test products and tools<br>o SP 1.1 Determine tool types<br>o SP 1.2 Compare the existing tools<br>o SP 1.3 Select specific tools and tool suppliers<br>o SP 1.4 Establish supplier agreements (in the case of commercial tools)<br>• SG 2 Manage test products and tools<br>o SP 2.1 Acquire, install and use test products (tools)<br>o SP 2.2 Maintain (get updates and support for) test products (tools)<br>o SP 2.3 If needed, develop in-house test tools |

## 4.3 REFINING THE PA "TEST SERVICES SUB-CONTRACT MANAGEMENT"

As discussed above, sub-contracting of software services are quite common, e.g., see the experience report in [48], and thus we needed to include that aspect in the customized model. There is a remotely-related PA in CMMI-SVC named "Supplier agreement management". We wanted to highlight the phrase "sub-contract" and "sub-contractors" in the PA title because that is the phrase most of industry folks are using in this context. We also refined the explanations slightly to make them more understandable for our industry partners, as shown in Table 6. The SP 2.1 "Monitor sub-contracted services" was added as an important activity to be done in this PA.

**Table 6- Refining the PA "Test services sub-contract management"**

| Original in CMMI-SVC: | How it was refined: |
|---|---|
| • SG 1 Establish supplier agreements<br>o SP 1.1 Determine acquisition type<br>o SP 1.2 Select suppliers<br>o SP 1.3 Establish supplier agreements<br>• SG 2 Satisfy supplier agreements<br>o SP 2.1 Execute the supplier agreement<br>o SP 2.2 Accept the acquired products<br>o SP 2.3 Ensure transition of products | • SG 1 Establish supplier agreements for test services to be sub-contracted<br>o SP 1.1 Determine acquisition type<br>o SP 1.2 Select suppliers<br>o SP 1.3 Establish supplier agreements<br>• SG 2 Manage sub-contracted services<br>o SP 2.1 Monitor sub-contracted services<br>o SP 2.2 Accept the sub-contracted services<br>o SP 2.3 Ensure transition of services from sub-contractor to client |

## 5 CASE STUDY: APPLICATION AND EVALUATION OF THE CUSTOMIZED MODEL

To assess the applicability and usefulness of CMMI-SVC, we designed and conducted a case study in two industrial settings. We report next the case study design, a description of the cases (contexts), and then the results. Finally, we discuss potential threats to the validity of our study and steps that we have taken to minimize or mitigate them.

### 5.1 CASE-STUDY DESIGN

#### 5.1.1 Goal, research questions and metrics

Stated using the goal template of the Goal, Question, Metric (GQM) approach [51], the goal of case study was to assess the applicability and usefulness of the CMMI-SVC model when it is applied in real industrial settings to assess the maturity of software testing services, from the point of view of stakeholders involved in software testing services (e.g., service provides and customers). Another side benefit of the case study was to provide a "trial" appraisal of testing services maturity in the companies under study (Section 5.1.2), which is a goal as reported in previous trial appraisals in industrial settings in the literature, e.g., SPICE (Software Process Improvement and Capability dEtermination) trials [52, 53].

As the above goal shows, the nature of our case study is 'exploratory' [54] in that our objective was to find out what is happening, to seek new insights, and to generate ideas and hypotheses for further research and follow-up improvements in test services for the companies. Based on the above goal, we posed one case-study question (CSQ) and three research questions (RQs):

- CSQ 1: What are the maturity levels of each of the industrial cases and what are the areas needing improvement in each case?
- RQ 1 (applicability): How applicable is the customized model? i.e., how easy it is to apply the model in industrial settings?
- RQ 2 (usefulness): To what extent do the stakeholders find the model useful for assessing the maturity and pinpointing improvement areas?
- RQ 3 (completeness and relevance): To what extent is the customized model complete and relevant to maturity appraisal of testing services? As discussed in Section 4.1, we customized CMMI-SVC to our context while ensuring its completeness and relevance in this context. (Completeness of the model in this context is whether it capture all the important aspects in the scope of maturity of testing services. Relevance in this context is whether all the factors and issues included in the model are relevant, i.e., should be assessed).

We raised the above CSQ to benefit the case-study companies and was not considered a research contribution. However, RQs 1-3 were research oriented. We planned to answer each of the above questions using both quantitative and qualitative metrics and then apply the triangulation technique [55] as discussed next.

For RQ 1 (model's applicability), we planned to ask the engineers involved in the industrial case studies, after the appraisals, to rank the extent to which that they found the model applicable in their company. We provided a 5-point Likert scale (as shown in For RQ 3 (model's completeness and relevance of the model), we planned to ask as many test engineers from our industry contacts as possible to assess completeness and relevance of the model based on the following 5-point Likert scale: (1-Very low, 2-Low, 3-Average, 4-High, and 5-Very high). Note that those test managers did not necessarily need to have used the model, but we would ask them to review the model in detail and rely on the assumption that if they were to use the model for appraisal. Thus, RQ 3 (completeness and relevance of the model) would provide a higher-level perspective compared to RQ 1 (model's applicability) and RQ 2 (model's usefulness).

Table 7) for that purpose. We also asked for their qualitative opinions (feedback) if the applicability was low.

For RQ 2 (model's usefulness), we planned to ask the engineers involved in the industrial case studies, after the appraisals, to rank the extent to which that they found the model useful for assessing the maturity and pinpointing improvement areas, both in quantitative and qualitative terms (as shown in For RQ 3 (model's completeness and relevance of the model), we planned to ask as many test engineers from our industry contacts as possible to assess completeness and relevance of the model based on the following 5-point Likert scale: (1-Very low, 2-Low, 3-Average, 4-High, and 5-Very high). Note that those test managers did not necessarily need to have used the model, but we would ask them to review the model in detail and rely on the assumption that if they were to use the model for appraisal. Thus, RQ 3 (completeness and relevance of the model) would provide a higher-level perspective compared to RQ 1 (model's applicability) and RQ 2 (model's usefulness).

Table 7). To get additional data points for RQ 2, we also gathered opinions of eight other test managers from our industry contacts, working in other companies, to assess the 'potential' usefulness of the model if they were to use the model in their contexts (those eight managers were not involved in the case study). For RQ 3 (completeness and relevance), we followed the same approach as RQ 2.

For RQ 3 (model's completeness and relevance of the model), we planned to ask as many test engineers from our industry contacts as possible to assess completeness and relevance of the model based on the following 5-point Likert scale: (1-Very low, 2-Low, 3-Average, 4-High, and 5-Very high). Note that those test managers did not necessarily need to have used the model, but we would ask them to review the model in detail and rely on the assumption that if they were to use the model for appraisal. Thus, RQ 3 (completeness and relevance of the model) would provide a higher-level perspective compared to RQ 1 (model's applicability) and RQ 2 (model's usefulness).

**Table 7-Metrics used to answer the study RQs**

| RQ | Metrics (Likert scales) | Qualitative feedback |
|---|---|---|
| RQ 1 (applicability) | 1. Very difficult<br>2. Difficult<br>3. Neutral<br>4. Easy<br>5. Very easy | If 'applicability < Neutral', please provide feedback |
| RQ 2 (usefulness) | 1. Not at all useful<br>2. Slightly useful<br>3. Moderately useful<br>4. Very useful<br>5. Extremely useful | If 'usefulness < Moderately useful', please provide feedback |
| RQ 3 (completeness and relevance) | 1. Very low<br>2. Low<br>3. Average | If 'completeness or relevance < Average', please provide feedback |
| | 4. High<br>5. Very high | |

### 5.1.2 A review of the cases (contexts): two industrial settings

Using our active industrial connections and partnerships, we conducted the case study in partnership with two industrial companies which have provided and are providing software testing services in the past and currently.

Note that we had access to a large pool of companies and industry contacts to choose from. To keep the effort levels manageable, we selected only two companies. In choosing the companies, we used both random and "stratified" sampling [56]. We divided the pool of companies in our network based on their ages, based on the rationale that, perhaps older companies would be more mature in test services, and then applied stratified sampling to choose one from each group (young versus old companies).

One company had an age of more than 10 years and the other company was less than five years old. For anonymity purposes, we refer to these two companies as C1 and C2. Both C1 and C2 provide software testing services actively in different domains in Turkey, e.g., government, defense and private sector. In terms of company sizes, both companies are Small and Medium-sized Enterprises (SMEs). In a given time, each company provides testing services to multiple clients in the scope of different testing projects. The projects are from different spectrums of software testing types, e.g., functional and non-functional testing. Thus, we believe that these two companies are good representatives of firms providing software testing services.

We, researchers, have had direct connections with several test engineers and managers in both C1 and C2 and have collaborated with them on other projects in the past. A test manager from C1 (TM1) and a test manager from C2 (TM2) agreed to be actively involved in this joint work for the case study purposes. Both TM1 and TM2 are regularly involved in the roles of service provider team leads. Thus, they were the most ideal subjects to be involved in the case study.

## 5.2 CASE-STUDY EXECUTION: APPLICATION OF THE MODEL

The execution of the case-study included application of the customized CMMI-SVC on each of the C1 and C2 cases, to extract the maturity appraisal results, and to answer the four RQs of the study. For this purpose, we setup regular meetings with TM1 and TM2 for the duration of several weeks. The iterative work started with a thorough introduction of the model by the researchers to TM1 and TM2 and several of their colleagues (test engineers working under their supervision) who were going to be also involved in the study.

To conduct the trial appraisals, we relied on the Standard CMMI Appraisal Method for Process Improvement (SCAMPI) [57], which suggested the following 6-step process: (1) Prepare participants, (2) Examine objective evidence, (3) Document objective evidence, (4) Verify objective evidence, (5) Validate preliminary findings, and (6) Generate appraisal results.

Staff in both organizations (including TM1 and TM2) were familiar with CMMI, since they were involved in conducting software process improvement (SPI) initiatives in the past based on CMMI for the government sector. This familiarity helped us in applying CMMI-SVC and increased the precision of gathered data.

As suggested by SCAMPI [57] (the 6-step process above), we as a group surveyed the organizations on whether there was evidence of the expected work products, practices or outcomes for each SP. For rank assessment of each SPs, we logged and included proper justifications/documentations for each item. Note that, again by following SCAMPI, we considered both "direct" (explicit) and "indirect" (implicit) artifacts for assessment of each SP. For example, for the SP "Analyze test service and customer needs", it could be that no explicit artifact (e.g., report or documentation) was available, but if there was even a suitable discussion in another document of the company, we considered the SP "implemented" (covered).

As per the above step # 3 of SCAMPI [57] ("Document objective evidence"), our intent was to "*create lasting records of the information gathered by identifying and then consolidating notes, transforming the data into records that document gaps in practice implementation or exemplary practice implementation*".

We stored all the model entities (SGs and SPs) in an Excel spreadsheet and we ranked them using the 5-point Likert scale, as suggested by SCAMPI [57]: (0) not implemented (NI), (1) partially implemented (PI), (2) largely implemented (LI), (4) fully implemented (FI), and (5) not yet (NY) or not applicable (NA). As suggested by SCAMPI [57], "*goal ratings were determined within each PA, which were then collectively used to determine aggregate ratings for the individual PAs*" and, consequently, the rating of the case company under study.

A snapshot of one of the assessment sheets, in which we recorded the "characterization" (score) for each SP and the corresponding evidence and supporting documentations/ justifications, is shown in Figure 7. For example, for SP 1.2 (discussing and negotiating service agreements), TM1 entered the comment that they always organize "*meetings with customers before bidding*".

| | A | B | C | D | E |
|---|---|---|---|---|---|
| 1 | | Process area | Specific goal and practices | NI, PI, LI, FI, N/A | Comments / justifications |
| 2 | 1 | Test service agreement establishment and delivery | · SG 1 Establish Service Agreements | | |
| 3 | | | o SP 1.1 Analyze test service and customer needs | LI | 1-2 weeks time for analyzing needs but because of security issues we can't analyze all details |
| 4 | | | o SP 1.2 Discus and negotiate elements of the Test Service Agreement (contract), e.g., basic details such as test requirements, duration/cost , and other issues that we have heard from our industry partners such as regression tests and when to stop | FI | Meetings with customers before bidding. |
| 5 | | | o SP 1.3 Establish the test Service Agreement (contract) | FI | We always have contracts. |
| 6 | | | · SG 2 Prepare for Service Delivery | | |
| 7 | | | o SP 2.1 Establish the Test Service Delivery Approach | FI | Usually described by the customer in contracts. |
| 8 | | | o SP 2.2 Prepare for Test Service Operations (mostly internal to the test team) | FI | |
| 9 | | | · SG 3 Deliver Services | | |
| 10 | | | o SP 3.1 Receive and Process Service Requests | FI | Requests are fixed at the beginning of the |
| 11 | | | o SP 3.2 Operate the Service System (including test staff, test processes and test activities) | FI | |
| 12 | | | o SP 3.3 Maintain the Test Service System (including test staff, test processes and test activities) | NA | We do not have a specific system. We work in project based contracts. |

**Figure 7- An excerpt from the spreadsheet in which we logged in detail the corresponding evidence and supporting documentations/ justifications for scoring each SP**

We should also note that we were aware of the critical success factors of Software Process Improvement (SPI), which was synthesized in a systematic review [58] as the following set: commitment, alignment with the business strategy and goals, training, communication, resources, skills, improvement management and staff involvement. While our case study only includes maturity appraisal and not maturity "improvement", we still did our best to maximize accuracy of our appraisal by considering the above critical success factors as much as possible. For example, we ensured getting the full commitment from participants in the case study. We also trained the participants, communicated regularly with all participants, and provided the required resources and skills.

## 5.3 Results

We address each of the study's CSQ and RQs below.

### 5.3.1 CSQ 1: Maturity levels of each industrial case and the areas needing improvement

After careful application of the model and case-study execution (Section 5.2), we recorded the characterization (score) for each SP along with the corresponding evidence and supporting justifications in a spreadsheet. Based on the individual SP rank values, we derived the consolidated trial appraisal using CMMI-SVC for each of the two industrial cases (as shown in Figure 8).

Since the type of maturity models can be either 'staged' or 'continuous', and since CMMI-SVC is a 'staged' models, we can see in Figure 8 that, in none of the two industrial cases, no single maturity level has been 'fully' implemented (having all green bars denoting 'FI'). This means that none of the two cases (firms) C1 or C2 met the full maturity in any of the four levels (even level #2). We discuss next some concrete examples of the SPs which received ratings of Fully Implemented (FI), Largely Implemented (LI), Partially Implemented (PI), Not Implemented (NI) or not applicable (NA) in each of the industrial cases.

As Figure 8 shows, for the case of C1, only 20 out of the 64 SPs in the level #2 are FI, e.g., SP 1.1 (Understand test requirements as documented in the test service agreement). Similarly, in the level #2, 20 SPs are also LI, e.g., SP 1.2 (Manage test requirements changes). Overall across all the four levels, for the case of C1, 92 out of the total 168 SPs (50.2%) are NI. By having a total of 99 SPs as NI (54.4%), the situation for C2 is slightly different. In summary, the appraisal shows the need for improvements in many of the areas in both test service companies.

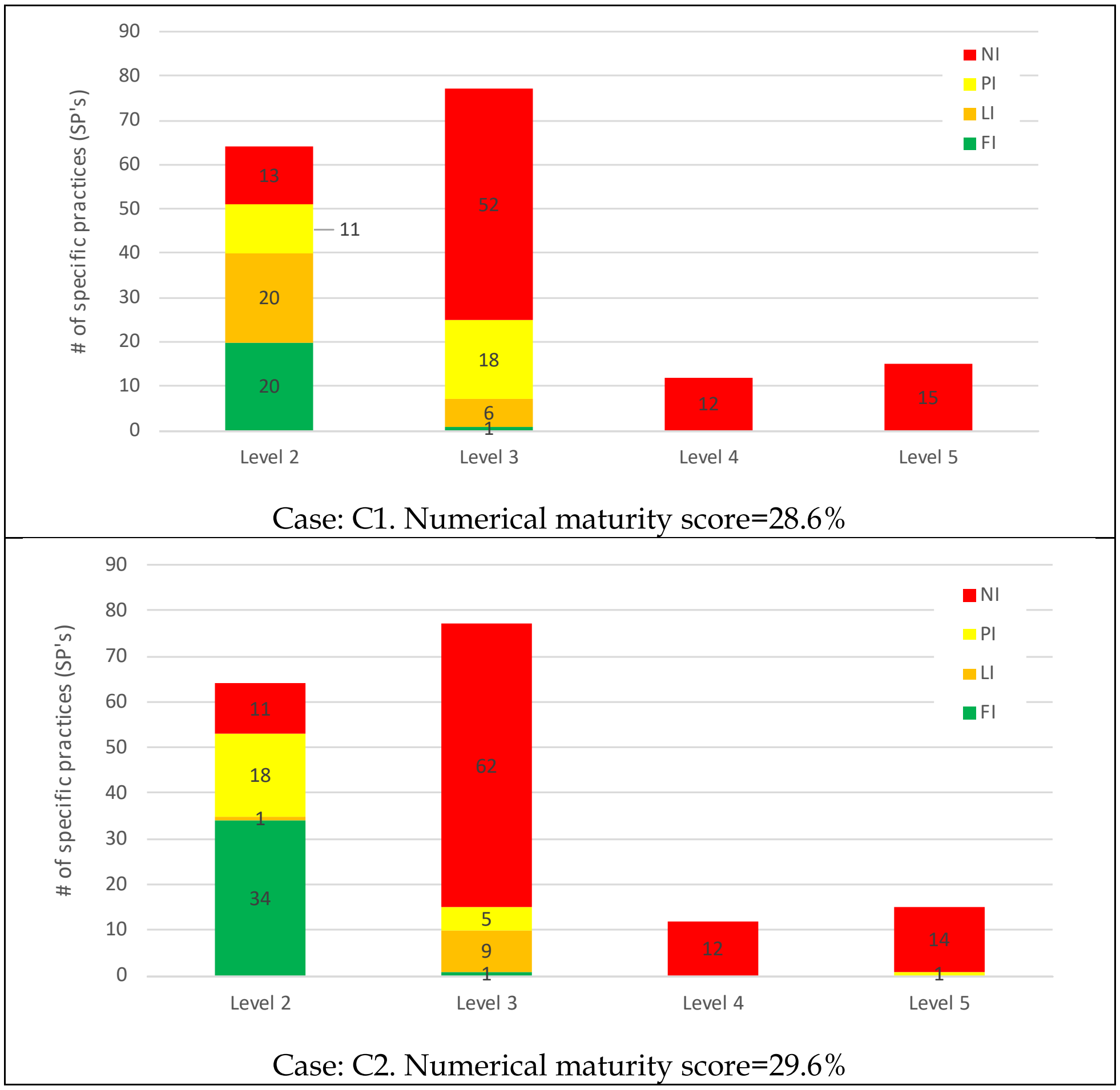

**Figure 8- Maturity appraisal results using CMMI-SVC for each of the two industrial cases**

We also wanted to numerically assess and compare the two cases using a single aggregate value. We used a weighting scheme to calculate a single percentage score for each of the cases. We assigned 3, 2, 1, and 0, respectively, to ranks FI, LI, PI, and NI (and N/A) and summed the values. Such a numerical assessment is also popular in the traditional software process improvement (SPI) literature, e.g., [59]. An assessment of all FI ranks for all the 168 SPs would yield a value of 504 (3*168). Based on the aggregate sum values (sum=144 for C1 and sum=149 for C2), the numerical scores for the two cases were 28.6% (C1) and 29.6% (C2).

Based on meetings and feedback from the companies, we have observed that the appraisal results have been helpful for the companies and the managers in knowing objectively where they (the quality of their testing services) stand and to pinpoint potential improvement areas. The ranks FI, LI, PI, NI and N/A explicitly showed to the test managers and the researchers the areas needing improvement in each case. Improvement activities have started in both the companies and are currently underway.

Recall from Section 5.2 that, while selecting the two cases for the empirical study, we used sampling based on company ages. Between C1 and C2, one had an age of more than 10 years and the other company was less than five years old. We hypothesized that, perhaps, an older company would be more mature in test services. But the data in Figure 8 did not support that hypothesis (their numerical maturity scores were so close). Note that since the number of cases were few (only two), we did not use formal hypothesis testing.

Last but not the least, we mentioned to our partner test managers that, to have a baseline (benchmark), it may be interesting to compare the trial appraisal results of the above two companies with official published CMMI-SVC appraisal results [60], and also TMMi-based appraisals of companies, e.g., [61, 62].

### 5.3.2 RQ 1: Applicability of the model

The first RQ was to see how applicable the CMMI-SVC model is, i.e., how easy it is to apply the model in industrial settings? Recall from Section 4.2 that a test manager from company C1 (TM1) and a test manager from C2 (TM2) were involved in the appraisals. Based a 5-point Likert scale of (1-Very difficult, 2-Difficult, 3-Neutral, 4-Easy, 5-Very easy), both TM1 and TM2 assessed the applicability of the model as '4-Easy'. They furthermore made the following comments. TM1 mentioned that: "*I personally think that the model should be used jointly by the service provider and customer. It would, in fact, be more beneficial that way. In many cases in my experience, service clients impose other sets of constraints which make us, as service providers, not apply many of the practices mentioned in the model. However, in general, we currently apply or think of applying many of the practices mentioned in this maturity model*" (the quotes in this section are translated from Turkish to English by the first author).

TM1 hoped that it would be nice if TMMI and CMMI-SVC would be like CMMI one day, in terms of adaptation and penetration in the industry in which many defense and government projects 'require' adoption of CMMI. If that occurs, TM1 believed that applying models such as CMMI-SVC would be much easier to do and also justify to upper-level management. TM1 further added that: "*I see three potential impediments for the applications of models such as CMMI-SVC: (1) support of upper-level management, (2) the culture of companies (both service providers and customers), and (3) dealing with and justifying the time/cost overhead spent on the maturity assessment using CMMI-SVC. If these potential impediments are addressed, applicability of the model would be high*". We thus observed that some of the concerns about CMMI-SVC adoption are not exclusive to it but to maturity models in general and we observed correspondence between the above-quoted challenges to some of the critical success factors in Software Process Improvement (SPI) in general [58], such as: commitment, alignment with the business strategy and goals, training, communication, resources, skills, improvement management and staff involvement. As discussed in Section 5.3, to maximize accuracy of our appraisal, we did our best by considering the above critical success factors as much as possible in the case study.

TM2 stated that: "*Our current (service delivery) methods are mostly overlapping with the process proposed by CMMI-SVC, but perhaps not too systematic in several PAs. For this reason, the applicability of the model for us was 'Easy'. Although applicability is easy, a maturity model brings extra effort and costs for us service providers. Services are generally outsourced by bidding processes. Adopting of maturity models such as CMMI-SVC may result in quality improvement but extra costs may cause a disadvantage in bidding processes. Therefore, the ideal case for applicability of proposed model is that customers would require the CMMI-SVC approval from *all* the service providers, so that test service providers who apply the model would not be in a disadvantageous position due to the costs of applying it*". Thus, as we observe, both TM1 and TM2 raised the cost and effort overheads of conducting maturity appraisal on software testing services using CMMI-SVC as a challenge. This is similar to other studies in the TMA and TPI literature in which maturity appraisal is seen as an "*extra project (efforts)*" [35, 63]. Raising and justifying the need to apply maturity appraisal on software testing services are issues that practitioners and researchers should focus on.

**5.3.3 RQ 2: Usefulness of the model**

We asked TM1 and TM2 to also assess the usefulness of the model based on the following 5-point Likert scale: (1-Not at all useful, 2-Slightly useful, 3-Moderately useful, 4-Very useful, 5-Extremely useful). TM1 and TM2 both assessed the usefulness of the model as '4-Very useful' and provided the following comments.

TM1 stated that: "*In my opinion, application of this model can provide a lot of benefits to other test service organizations, as it did for our case. In the past, we have conducted and successfully delivered a number of test service projects, in which, most of our practices matched the process areas (PAs) exactly mentioned in CMMI-SVC! Now that I think about those experiences, I see that for new companies, CMMI-SVC will help them jump-start their maturity without having to go through the (painful) 'try-and-error' leaning curves, which we went through! Also I see that application of this model can clearly increase the quality of the work practices*".

TM2 stated that: "*For our company, the model has shown to be very useful. A test service is much different than a software development service. There are several reasons for this difference but the most important is that there are no obvious prescriptions/rules for fuzzy issues such as 'When to stop testing' and 'How much regression testing to do'. This perspective brings many complexities to define the scope of a test service. A defined maturity model such as CMMI-SVC can eliminate the complexity and identify a way in which customer and provider can walk through together without problems*".

Apart from asking TM1 and TM2 to assess the usefulness of the model, as discussed in Section 5.1, to enrich our analysis for RQ 2, we also contacted eight other test managers from the list of our industry contacts (whom with we had collaborated in our past projects), working in other companies, to assess the 'potential' usefulness of the model if they were to use the model in their contexts. Their opinions are shown in Figure 9. As we can see, their general opinion is that the model is potentially useful for them, even if they have not yet used it for assessing maturity in their contexts. One of those test managers, who voted 'Very useful', stated that "*I am really glad to have seen this model. I am going to try it in our upcoming test service projects*". One test manager voted 'Slightly useful' and expressed that "*…to be able to assess the usefulness of CMMI-SVC, I should use it in our projects first*".

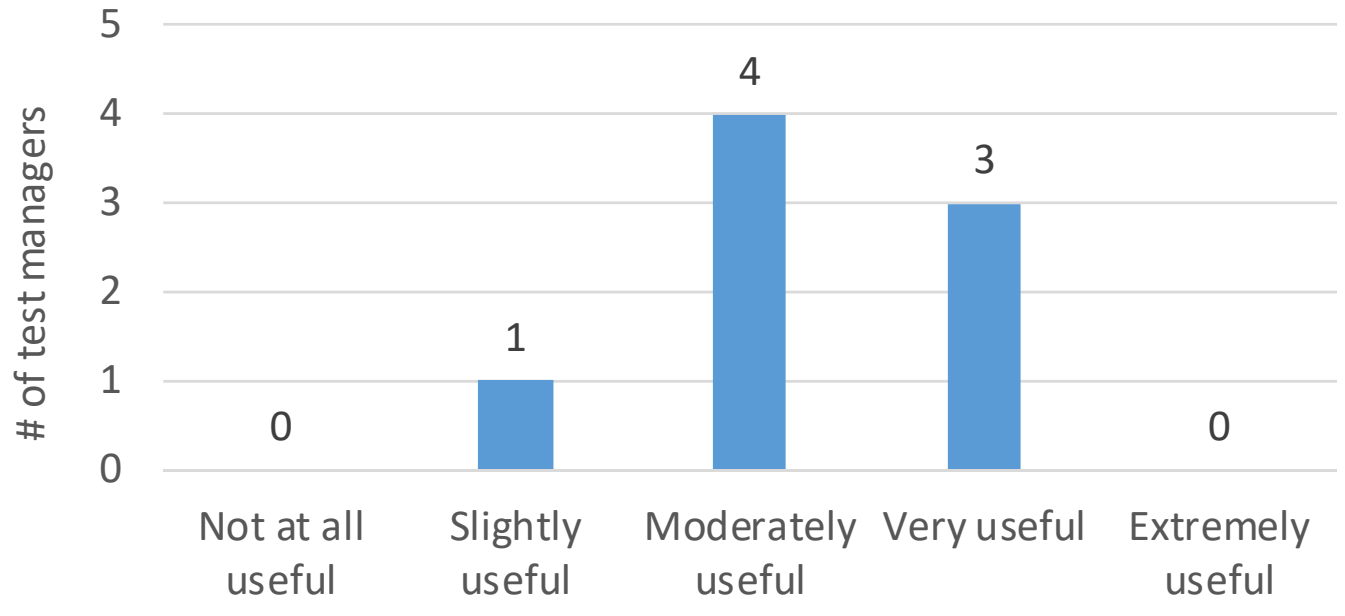

**Figure 9- Opinions of eight test managers on 'potential' usefulness of CMMI-SVC**

**5.3.4 RQ 3: Completeness and relevance of the model**

For our last RQ, we asked eight test managers from our industry contacts to assess completeness and relevance of the model based on the following 5-point Likert scale: (1-Very low, 2-Low, 3-Average, 4-High, and 5-Very high). Figure 10 shows the summary of the opinion survey.

Opinions shown in Figure 10 depict a general agreement on high completeness and relevance of the model. We were expecting such an outcome from the opinion survey, since as discussed in Section 4, while customizing the model, we considered completeness and relevance explicitly as two important criteria. This was done to ensure that the customized model would meet the needs of the software testing industry. We iteratively customized the model in several iterations by getting opinions from three expert (senior) test engineers and managers (outside the authors list).

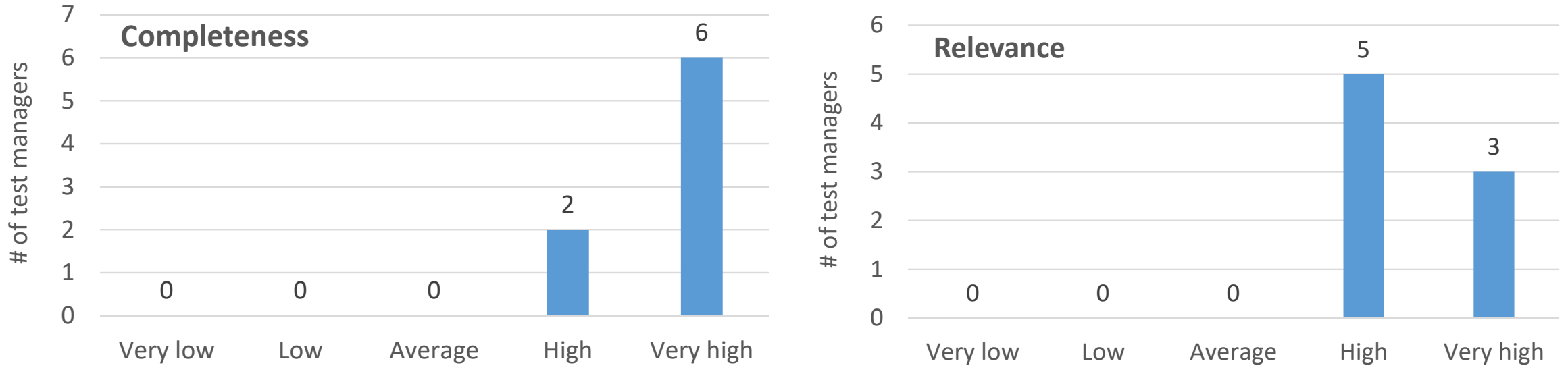

**Figure 10- Opinions of eight test managers on completeness and relevance of the model**

## 5.4 Limitations and potential threats to validity

In design and execution of the case study, we considered potential threats to the validity of our study and discuss next the steps that we have taken to minimize or mitigate them. The threats are discussed in the context of the four types of threats to validity based on a standard checklist for validity threats [64]: internal validity, construct validity, conclusion validity and external validity.

<u>Internal validity:</u> Internal validity is a property of scientific studies which reflects the extent to which a causal conclusion based on a study and the extracted data is warranted [64]. A threat to internal validity in this study lies in the selection bias (i.e., randomness of the companies and test managers who participated in our case studies). As discussed in Section 5.2,to keep our effort levels manageable, we selected only two companies but are planning to conduct more trial appraisals in the future. We also ensured that we chose test managers from different industrial domains. Also since we had access to a large pool of companies and industry contacts to choose from, we choose the companies randomly. Thus we believe they are reasonable representative of companies offering testing services.

<u>Construct validity:</u> Construct validities are concerned with the extent to which the objects of study truly represents theory behind the study [64]. In other words, the issue relates to whether the model actually helps measuring the maturity of software testing services. As discussed in Section 5, we considered two important criteria while assessing the model: completeness and relevance. By giving importance to completeness, we assessed whether the model captures the important aspects in the scope of maturity of testing services, as needed in real-world projects. By giving importance to relevance, we ensured that each of the factors and issues included in the model was relevant, i.e., should be assessed. By getting experts' opinions, results of RQ 3 also confirmed that indeed, the model is complete and relevant to a high degree. One other aspect under construct validity was "face" validity [65], i.e., does the model and our approach "look like" a suitable measure of the interest? We assessed face validity w.r.t. these questions: (1) Does the model and our approach "look like" a suitable measure of the desired construct to a member of the target population (a typical test manager)? and (2) Did test manager recognize the type of information they were responding to? Based on the literature, we were aware possible advantages of face validity, e.g., if the respondent knows what information we were looking for when conducting the trial appraisal, they could use that "context" to help interpret the questions and provide more useful, accurate answers. We believe that since the results of RQ 1 showed high applicability of the model and results of RQ 3 showed high completeness and relevance of the model, we have been able to achieve high face validity.

Another aspect under construct validity was "content" validity [65], i.e., does the model and our approach contain items from the desired "content domain"? Content validity is usually "assured" by the "*informed item selections made by experts in the domain*" [66]. Since we closely followed the CMMI SCAMPI guideline [57], and its step #1 was to "Prepare participants", we ensured that the participants made "informed" selections and provided informed evidence and characterizations for each SP of each PA. Thus, we assured to achieve high content validity.

Conclusion validity: Conclusion validity of a study deals with whether correct conclusions are reached through rigorous and repeatable treatment [64]. In our case study, we attempted to assess the applicability and usefulness of the CMMI-SVC model when it is applied in real industrial settings, by raising two RQs. For each RQ, we attempted to reduce the bias by seeking support from the 5-point Likert scale data, gathered via the opinion survey. We answered each of the RQs using both quantitative and qualitative metrics. Thus, all the conclusions that we drew in this study are strictly traceable to data.

External validity: External validity is concerned with the extent to which the results of this study can be generalized [64]. All the efforts were made to minimize the selection bias, which is an important factor for both internal and external validity. Although we only applied the model in two industrial settings, we believe they are typical test-service companies (as per our long track record in working with many other firms in the past). We thus believe the two cases are suitable representatives of such firms. However, it is obvious that maturity appraisal of other test-service companies would provide different results. Last but not the least, it is needless to say that the CMMI-SVC model is for maturity appraisal of testing services and shall not be used for other purposes.

## 6 Conclusions and future works

Based on a real industrial need, we chose, customized and empirically applied the 'CMMI for Services' (CMMI-SVC) model for assessing the maturity of software testing services. To assess the applicability and usefulness of the model, we evaluated the model in two industrial settings by applying it in two companies who provide software testing services in Turkey. The quantitative and qualitative results of the case study have shown that the customized model has been indeed useful and helpful for both companies and their test management teams by helping them objectively assess the maturity of their testing services and also to pinpoint potential improvement areas.

Note that while we reported two successful trial appraisals of testing services maturity using CMMI-SVC in this paper, we, by no mean, propose (or want to convey) that appraisal w.r.t. CMMI-SVC should be necessary for all providers of testing services. As it is widely known, it took years for the main CMMI model to get established and was mainly due to strong endorsement (and sometimes enforcement) of clients of software projects, e.g., the US government contracts often require bidders to have achieved some level of CMMI [67]. Thus, requiring maturity appraisals using CMMI-SVC or any other model is an issue which shall be ultimately decided by the entity receiving the test service (client) and the software industry in general.

Our future work directions include the followings: (1) further empirical application of the model by applying it in more industrial settings and to further improve the model; (2) conducting empirical studies on the relationship of the maturity score as assessed by this model and the quality of testing services offered, i.e., does a high maturity score necessarily translate to high-quality testing services?; (3) studying the correlations among the TMMI, CMMI and CMMI-SVC ratings of a given industrial context; and (4) once we hopefully conduct a few more empirical studies and trial appraisals like the one reported in this paper, we intend to conduct cross-case meta analyses such as the one reported for SPICE trials [52], in which we would look into the following aspects: (i) investigation into reasons for performing test maturity appraisal, (ii) evaluation of the internal consistency of the dimensions in the CMMI-SVC model when used for testing services, (iii) using interrater agreement as a measure of the reliability of appraisals, (iv) evaluation of the predictive validity of process capability, and (v) identification of factors influencing assessor effort.

## Acknowledgements

The authors would like to thank the test managers and test engineers in several different companies for their involvement in the industrial case studies. Vahid Garousi was partially supported by several internal grants provided by the Scientific and Technological Research Council of Turkey (TÜBİTAK) via grant #115E805.

## References

[1] T. Britton, L. Jeng, G. Carver, P. Cheak, and T. Katzenellenbogen, "Reversible Debugging Software," *University of Cambridge, Judge Business School, Tehnical Report*, 2013.

[2] V. Garousi, A. Coşkunçay, A. B. Can, and O. Demirörs, "A survey of software engineering practices in Turkey," *Journal of Systems and Software,* vol. 108, pp. 148-177, 2015.

[3] V. Garousi and J. Zhi, "A survey of software testing practices in Canada," *Journal of Systems and Software,* vol. 86, no. 5, pp. 1354-1376, 2013.

[4] V. Garousi and T. Varma, "A replicated survey of software testing practices in the Canadian province of Alberta: what has changed from 2004 to 2009?," *Journal of Systems and Software,* vol. 83, no. 11, pp. 2251-2262, 2010.

[5] O. Taipale and K. Smolander, "Improving software testing by observing practice," in *Proceedings of the 2006 ACM/IEEE international symposium on Empirical software engineering*, 2006, pp. 262-271: ACM.

[6] V. Garousi, A. Coşkunçay, A. B. Can, and O. Demirörs, "A survey of software testing practices in Turkey (Türkiye'deki yazılım test uygulamaları anketi)," in *Proceedings of the Turkish National Software Engineering Symposium "Ulusal Yazılım Mühendisliği Sempozyumu" (UYMS)*, 2013.

[7] S. Zogaj, U. Bretschneider, and J. M. Leimeister, "Managing crowdsourced software testing: a case study based insight on the challenges of a crowdsourcing intermediary," *Journal of Business Economics,* journal article vol. 84, no. 3, pp. 375-405, 2014.

[8] R. S. Poston, J. C. Simon, and R. Jain, "Client Communication Practices in Managing Relationships with Offshore Vendors of Software Testing Services," *Communications of the Association for Information Systems,* vol. 27, pp. 129-148, 2010.

[9] H. Shah, M. J. Harrold, and S. Sinha, "Global software testing under deadline pressure: Vendor-side experiences," *Information and Software Technology,* vol. 56, no. 1, pp. 6-19, 2014.

[10] J. Gao, X. Bai, W. T. Tsai, and T. Uehara, "Testing as a Service (TaaS) on Clouds," in *IEEE International Symposium on Service-Oriented System Engineering*, 2013, pp. 212-223.

[11] L. Yu, L. Zhang, H. Xiang, Y. Su, W. Zhao, and J. Zhu, "A Framework of Testing as a Service," in *International Conference on Management and Service Science*, 2009, pp. 1-4.

[12] Guru99, "Software Testing as a Service (TaaS)," *http://www.guru99.com/what-is-testing-as-a-service-taas.html*, Last accessed: Aug. 2017

[13] E. v. Veenendaal and B. Wells, *Test Maturity Model integration (TMMi): Guidelines for Test Process Improvement*. Uitgeverij Tutein Nolthenius, 2012.

[14] TMMI Foundation, "TMMI specification (reference model), release 1.0," *http://www.tmmi.org/pdf/TMMi.Framework.pdf*, Last accessed: Aug. 2017.

[15] T. Koomen and M. Pol, *Test Process Improvement: A Practical Step-by-step Guide to Structured Testing*. Addison-Wesley, 1999.

[16] V. Garousi, M. Felderer, and T. Hacaloğlu, "Software test maturity assessment and test process improvement: A multivocal literature review," *Information and Software Technology,* vol. 85, pp. 16–42, 2017.

[17] W. Afzal, S. Alone, K. Glocksien, and R. Torkar, "Software test process improvement approaches: A systematic literature review and an industrial case study," *Journal of Systems and Software,* vol. 111, pp. 1-33, 2016.

[18] V. Garousi and K. Herkiloğlu, "Selecting the right topics for industry-academia collaborations in software testing: an experience report," in *IEEE International Conference on Software Testing, Verification, and Validation*, 2016, pp. 213-222.

[19] V. Garousi, M. M. Eskandar, and K. Herkiloğlu, "Industry-academia collaborations in software testing: experience and success stories from Canada and Turkey," *Software Quality Journal, special issue on Industry Academia Collaborations in Software Testing*, pp. 1-53, 2016.

[20] M. E. Coşkun, M. M. Ceylan, K. Yiğitözu, and V. Garousi, "A tool for automated inspection of software design documents and its empirical evaluation in an aviation industry setting," in *Proceedings of the International Workshop on Testing: Academic and Industrial Conference - Practice and Research Techniques (TAIC PART)*, 2016, pp. 118-128.

[21] G. Urul, V. Garousi, and G. Urul, "Test Automation for Embedded Real-time Software: An Approach and Experience Report in the Turkish Industry," in *Proceedings of the Turkish National Software Engineering Symposium "Ulusal Yazılım Mühendisliği Sempozyumu" (UYMS)*, 2014.

[22] S. A. Jolly, V. Garousi, and M. M. Eskandar, "Automated Unit Testing of a SCADA Control Software: An Industrial Case Study based on Action Research," in *IEEE International Conference on Software Testing, Verification and Validation (ICST)*, 2012, pp. 400-409.

[23] C. P. Team, "CMMI for Services, version 1.3," *CMU technical report#: CMU/SEI-2010-TR-034*, 2010.

[24] E. Forrester, B. Buteau, and S. Shrum, *CMMI for Services: Guidelines for Superior Service*. Pearson Education, 2011.

[25] P. S. M. d. Santos and G. H. Travassos, "Action research use in software engineering: An initial survey," presented at the Proceedings of the 2009 3rd International Symposium on Empirical Software Engineering and Measurement, 2009.

[26] D. E. Avison, F. Lau, M. D. Myers, and P. A. Nielsen, "Action research," *Communications of the ACM*, vol. 42, no. 1, pp. 94-97, 1999.

[27] T. Gorschek, C. Wohlin, P. Carre, and S. Larsson, "A Model for Technology Transfer in Practice," *IEEE Software*, vol. 23, no. 6, pp. 88-95, 2006.

[28] V. Garousi, K. Petersen, and B. Özkan, "Challenges and best practices in industry-academia collaborations in software engineering: a systematic literature review," *Information and Software Technology*, vol. 79, pp. 106–127, 2016.

[29] C. G. v. Wangenheim, J. Carlo, J. C. R. Hauck, and L. A. V. Wangenheim, "Systematic Literature Review of Software Process Capability/Maturity Models," in *Proceedings of International Conference on Software Process. Improvement and Capability dEtermination (SPICE)*, 2010.

[30] I. Menken, *ISO/IEC 20000 Certification and Implementation Guide*, Second Edition ed. Emereo Pty Limited, 2009.

[31] Jan van Bon, *ITIL® 2011 Edition - A Pocket Guide*. Van Haren, 2011.

[32] Research and Markets Co., "Global Software Testing Services Market 2016-2020," *http://www.researchandmarkets.com/research/gj9hct/global_software*, 2015, Last accessed: Aug. 2017.

[33] Infosys Co., "Infosys Validation Solutions-Case Studies," *https://www.infosys.com/IT-services/validation-solutions/case-studies/*, Last accessed: Aug. 2017.

[34] Y. Lu and T. Käkölä, "A Dynamic Life-cycle Model for the Provisioning of Software Testing Services: Experiences from A Case Study in the Chinese ICT Sourcing Market," in *European Conference on Information Systems*, 2011.

[35] V. Garousi, M. Felderer, and T. Hacaloğlu, "What we know about software test maturity and test process improvement," *IEEE Software, In press*, 2017.

[36] M. D. Karr, "The Unit Test Maturity Model," *http://davidmichaelkarr.blogspot.com.tr/2013/01/the-unit-test-maturity-model.html*, 2013, Last accessed: Aug. 2017.

[37] G. Hongying and Y. Cheng, "A customizable agile software Quality Assurance model," in *Information Science and Service Science (NISS), 2011 5th International Conference on New Trends in*, 2011, vol. 2, pp. 382-387: IEEE.

[38] M. H. Krause, "A Maturity Model for Automated Software Testing," *http://www.mddionline.com/article/software-maturity-model-automated-software-testing*, 1994, Last accessed: Aug. 2017.

[39] E. Jung, "A test process improvement model for embedded software developments," in *Quality Software, 2009. QSIC'09. 9th International Conference on*, 2009, pp. 432-437: IEEE.

[40] S. Ronen- Harel, "ATMM Agile Testing Maturity Model," *http://www.slideshare.net/AgileSparks/atmm-practical-view*, 2010, Last accessed: Aug. 2017.

[41] T. Schweigert, A. Nehfort, and M. Ekssir-Monfared, "The Feature Set of TestSPICE 3.0," in *Systems, Software and Services Process Improvement*: Springer, 2014, pp. 309-316.

[42] S. Reid, "The Personal Test Maturity Matrix," *CAST 2006: Influencing the Practice June 5th-7th, 2006–Indianapolis*, p. 133, 2006.

[43] I. Burnstein, A. Homyen, R. Grom, and C. R. Carlson, "A Model to Assess Testing Process Maturity," *Crosstalk: The Journal of Defense Software Engineering*, vol. 11, 1998.

[44] A. v. Ewijk, B. Linker, M. v. Oosterwijk, and B. Visser, *TPI next: business driven test process improvement*. Kleine Uil, 2013.

[45] M. B. Chrissis, M. Konrad, and S. Shrum, *CMMI: Guidelines for Process Integration and Product Improvement*. Addison-Wesley, 2003.

[46] M. Tarnowski, "Capability Maturity Model Integration (CMMI)," *http://www.plays-in-business.com/cmmi-capability-maturity-model-integration/*, 2011, Last accessed: Aug. 2017.

[47] Lamri Corp., "The Compelling Case For CMMI-SVC: CMMI-SVC, ITIL & ISO20000 demystified," *http://www.lamri.com/files/resources/20KDemystified.pdf*, Last accessed: April 2016.

[48] J. Rudzki, T. Systä, and K. Mustonen, "Subcontracting Processes in Software Service Organisations - An Experience Report," in *Trustworthy Software Development Processes: International Conference on Software Process, ICSP 2009 Vancouver, Canada, May 16-17, 2009 Proceedings*, Q. Wang, V. Garousi, R. Madachy, and D. Pfahl, Eds. Berlin, Heidelberg: Springer Berlin Heidelberg, 2009, pp. 224-235.

[49] C. A. Cianfrani and J. E. J. West, *ISO 9001:2015 Explained*. ASQ Quality Press, 2015.

[50] V. Garousi, S. Arkan, G. Urul, and Ç. M. Karapıçak, "Maturity levels of the customized CMMI-SVC for testing services and their process areas," *Zenodo. http://doi.org/10.5281/zenodo.840039*, Last accessed: Aug. 2017.

[51] V. R. Basili, "Software modeling and measurement: the Goal/Question/Metric paradigm," Technical Report, University of Maryland at College Park1992.

[52] H.-W. Jung, R. Hunter, D. R. Goldenson, and K. El-Emam, "Findings from Phase 2 of the SPICE trials," *Software Process: Improvement and Practice*, vol. 6, no. 4, pp. 205-242, 2001.

[53] Y. Wang, A. Dorling, H. Wickberg, and G. King, "Experience in comparative process assessment with multi-process-models," in *Proceedings 25th EUROMICRO Conference. Informatics: Theory and Practice for the New Millennium*, 1999, vol. 2, pp. 268-273 vol.2.

[54] P. Runeson and M. Höst, "Guidelines for conducting and reporting case study research in software engineering," (in English), *Empirical Software Engineering*, vol. 14, no. 2, pp. 131-164, 2009.

[55] U. Flick, *How to combine multiple research options: Practical Triangulation*. SAGE, 2009.

[56] T. R. Lunsford and B. R. Lunsford, "The Research Sample, Part I: Sampling," *J. Prosthetics and Orthotics*, vol. 7, pp. 105-112, 1995.

[57] M. B. Chrissis, M. Konrad, and S. Shrum, *CMMI: Guidelines for Process Integration and Product Improvement*. Addison-Wesley, 2007.

[58] S. Bayona, J. A. Calvo-Manzano, and T. San Feliu, "Critical Success Factors in Software Process Improvement: A Systematic Review," in *International Conference on Software Process Improvement and Capability Determination*, 2012, pp. 1-12.

[59] A. April and C. Laporte, "An Overview of Software Engineering Process and Its Improvement," in *Encyclopedia of Information Science and Technology*Second ed.: IGI Global, 2009.

[60] CMMI® Institute, "Published CMMI® Appraisal Results," https://sas.cmmiinstitute.com/pars/pars.aspx, 2012, Last accessed: Aug. 2017.

[61] K. G. Camargo, F. C. Ferrari, and S. C. Fabbri, "Characterising the state of the practice in software testing through a TMMi-based process," *Journal of Software Engineering Research and Development*, vol. 3, no. 1, pp. 1-24, 2015.

[62] K. Rungi and R. Matulevičius, "Empirical Analysis of the Test Maturity Model Integration (TMMi)," in *Information and Software Technologies*, vol. 403, T. Skersys, R. Butleris, and R. Butkiene, Eds. (Communications in Computer and Information Science: Springer Berlin Heidelberg, 2013, pp. 376-391.

[63] J. García, A. de Amescua, M. Velasco, and A. Sanz, "Ten factors that impede improvement of verification and validation processes in software intensive organizations," *Software Process: Improvement and Practice*, vol. 13, no. 4, pp. 335-343, 2008.

[64] C. Wohlin, P. Runeson, M. Höst, M. C. Ohlsson, B. Regnell, and A. Wesslén, *Experimentation in Software Engineering: An Introduction*. Kluwer Academic Publishers, 2000.

[65] R. J. Gregory, *Psychological Testing: History, Principles and Applications*. Pearson Education, 2013.

[66] C. Garbin, "Course materials for Psychology 451/851 (Multivariate Research Design & Data Analysis)," http://psych.unl.edu/psycrs/451/e2/fccvalidity.pdf, Last accessed: Aug. 2017.

[67] T. Bollinger and C. McGowan, "A Critical Look at Software Capability Evaluations: An Update," *IEEE Software*, vol. 26, no. 5, pp. 80-83, 2009.

## APPENDIX- ACRONYMS USED IN THE PAPER

| | |
|---|---|
| CAM | Capacity and Availability Management |
| CMM | Capability Maturity Model |
| CMMI | Capability Maturity Model Integration |
| CMMI-ACQ | CMMI for Acquisition |
| CMMI-DEV | CMMI for Development |
| CMMI-SVC | CMMI for Services |
| CSQ | Case-study question |
| IAC | Industry-Academia Collaboration |
| IRP | Incident resolution and prevention |
| ITIL | Information Technology Infrastructure Library |
| KA | Key Area |
| KPA | Key Performance Areas |
| MLR | Multivocal Literature Review |
| PA | Process Area |
| RD | Requirements development |
| RQ | Research question |
| SCAMPI | Standard CMMI Appraisal Method for Process Improvement |
| SCON | Service continuity |
| SD | Service delivery |
| SG | Specific Goal |
| SLR | Systematic Literature Review |
| SP | Specific Practice |
| SPICE | Software Process Improvement and Capability dEtermination |
| SSD | Service system development |
| SSM | Strategic service management |
| SST | Service system transition |
| SUT | Software Under Test |
| TaaS | Testing as a Service |
| TMA | Test Maturity Assessment |
| TMMi | Test Maturity Model integration |
| TPI | Test Process Improvement |
| TS | Technical solution |